\documentclass[dvips]{article}
\usepackage{icrctc07}

\newcommand{\lesssim}{\lower.5ex\hbox{$\; \buildrel < \over\sim \;$}}
\newcommand{\gtrsim}{\lower.5ex\hbox{$\; \buildrel > \over\sim \;$}}
\newcommand{\e} {\epsilon}
\newcommand{\ep} {\epsilon^\prime}
\newcommand{\g} {\gamma}
\newcommand{\gp} {\gamma^\prime}
\newcommand{\G} {\Gamma}
\newcommand{\tp} {t^\prime}

\newcommand{\sT}{\sigma_{\rm T}}

\def\kms{\mbox{$\;$km s$^{-1}$}}     
\def\ergs{\mbox{$\;$ergs}}
\def\ergss{\mbox{$\;$ergs s$^{-1}$}}

\title{On Gamma Ray Burst and Blazar AGN
 Origins of the Ultra-High Energy Cosmic Rays
in Light of First Results from Auger
}
\shorttitle{GRB and Blazar Origins of the UHECRs}
\authors{Charles D.\ Dermer$^{1}$}
\shortauthors{C.\ D.\ Dermer}
\afiliations{$^1$Space Science Division, Code 7653, U.S. Naval Research Laboratory, Washington, D.C. 20375-5352 USA}
\email{dermer@ssd5.nrl.navy.mil}

\abstract{The discoveries of the GZK cutoff with the HiRes 
and Auger Observatories and the discovery by Auger of 
clustering  of $\gtrsim 60$ EeV ultra-high
energy cosmic rays (UHECRs) towards nearby ($\lesssim 75$ Mpc) AGNs
along the supergalactic plane establishes the astrophysical
origin of the UHECRs. The likely sources of the UHECRs are gamma-ray
bursts and radio-loud AGNs because: (1) they are extragalactic; (2)
they are sufficiently powerful; (3) acceleration to ultra-high
energies can be achieved in their relativistic ejecta; (4) anomalous
X-ray and $\gamma$-ray features can be explained by nonthermal hadron
acceleration in relativistic blast waves; and (5) sources reside
within the GZK radius. Two arguments
for acceleration to UHE are presented, and limits on UHECR ion
acceleration are set.
UHECR ions are shown to be able to survive without 
photodisintegrating
while passing through the AGN scattered radiation field, even if launched
deep in the broad line region.  UHECR
injection throughout cosmic time fits the measured energy spectrum of
UHECRs, at least for protons.  Local UHECR proton and ion 
interaction and energy-loss mean free paths are calculated using an
empirical fit to the extragalactic background light (EBL) at IR and
optical energies. Minimum intergalactic magnetic (IGM) fields 
$\approx 10^{-11}$ G are derived from
 clustering assuming specific source origins, e.g., Cen A, 
nearby AGNs, or GRBs for the super-GZK CRs seen with Auger. 
Besides distinct cosmic-ray induced $\gamma$-ray
signatures that should be observed with the {\it Gamma ray Large Area
Space Telescope (GLAST)}, source and GZK neutrino detections and the
arrival distribution of UHECR in direction and time can finally 
decide the sources of cosmic rays at the highest energies. }

\email{charles.dermer@nrl.navy.mil}

\begin{document}
\maketitle

\section{Introduction}
A high-significance steepening in the UHECR spectrum at energy $E\cong
10^{19.6}$ eV was reported earlier this year by the HiRes
collaboration \cite{hires07}, and here at the 2007 M\'erida ICRC based
on observations taken with the Auger Observatory \cite{yam07}.  This
result confirms the prediction of the Greisen-Zatsepin-Kuzmin (GZK)
cutoff \cite{gre66,zk66} in the UHECR spectrum due to photohadronic
interactions of UHECRs with photons of the cosmic microwave background
radiation (CMBR), and favors astrophysical bottom-up vs. particle
physics top-down scenarios for the UHECRs, provided that sources are
found within the GZK radius.

At the same time, Auger data shows \cite{ung07} evidence for mixed
composition with substantial ion content in UHECRs with energies as
high as a $few \times 10^{19}$ eV, based on studies of the depth of
shower maxima. With hybrid fluorescence detectors and shower
counters, Auger provides the strongest evidence yet for metals in the
UHECRs, possibly with mean atomic mass $\langle A \rangle \sim 8$ --
26, significantly different from pure proton and pure Fe composition.
This result depends on the accuracy of the nuclear
interaction physics used to model showers, but points to the
importance of nuclei in the UHECRs, and the meaning of this for GZK
physics \cite{nw00,ta04,alo07}.

The GZK radius of a proton with energy $E= E_{par} = 10^{20}E_{20}$ 
eV can be estimated by noting
that the product of the cross section and the inelasticity in a
pion-producing reaction is $K_{\phi\pi}\sigma_{\phi\pi} \approx 70~
\mu$b \cite{ad01}, so that the photopion energy-loss pathlength
$r_{\phi\pi}(E)$ is given by $n_{ph}(E)(K_{\phi\pi}\sigma_{\phi\pi})
r_{\phi\pi}(E) = 1$, where $n_{ph}(E)$ is the $E$-dependent number
density of photons above the pion-production threshold.

There are $\approx 400$ CMBR photons per cm$^3$ at the present epoch,
and ions with Lorentz factor $\gamma=E/Am_pc^2$ satisfying
$\gamma(h\nu/m_ec^2)\gtrsim 2m_\pi/m_e \cong 500$ will interact with
most of the photons of the CMBR. The mean dimensionless photon energy
in the CMBR is $h\langle \nu\rangle/m_ec^2 \cong 2.70 k_{\rm
B}T_{CMBR}/m_ec^2\cong 1.3\times 10^{-9}$, noting that $T_{CMBR} =
2.72$ K at the present epoch; hence UHECR protons with Lorentz factor
$\gamma\gtrsim 4\times 10^{11}$, or $E\gtrsim 4\times 10^{20}$ eV will
have a photo-pion energy-loss pathlength $r_{\phi\pi}(E\gtrsim 4\times
10^{20}$ eV) $\approx 12$ Mpc.  By considering the number of photons
above threshold in a blackbody distribution, the energy-loss mean free
path of an UHECR proton with energy $E$ is found to
be
\begin{equation}
r_{\phi\pi}(E_{20}) \cong {13.7 \exp[4/E_{20}] \over [1 +
4/E_{20}]}\;{\rm Mpc}\;.
\label{eq1}
\end{equation}
Figure 1 shows the photopion energy-loss pathlength from eq.\
(\ref{eq1}) for cosmic-ray protons interacting with CMBR photons, in
comparison with a more detailed calculation \cite{sta00}.

Auger results show a mixed UHECR composition extending from $\approx
4.5\times 10^{17}$ eV -- $4.5\times 10^{19}$ eV, with an indication of
increased average mass at the highest energy datum \cite{ung07}. The
ionic content of the UHECR provides new information to understand
UHECR source properties, especially if the distribution of atomic mass
$A$ can be obtained with better statistics.

Figure 1 shows expansion, photohadronic, and total energy-loss 
mean free paths
(MFPs) for p and Fe on the CMBR. 
Calculations
of photopair losses follow
\cite{czs92}.  
 The photopion
loss calculation can be extended to ions of atomic mass $A$ and charge $Z$ by
assuming that the photopion cross-section times inelasticity 
is at most $= A^{2/3}
K_{\phi\pi}\sigma_{\phi\pi}$ above threshold.
The photopion energy-loss rate of Fe with this assumption for the
product of the inelasticity and cross section is shown in Figure 1.
Because of the greater mass of Fe than p, far fewer photons are
available for photopion interactions than with protons carrying the
same energy, so that the corresponding GZK photopion radius is much
larger.  

Ions are in addition subject to losses due to photonuclear
interactions that can break up nuclei (for example, via the giant
dipole resonance).  The photodisintegration loss length for Fe in the
CMBR at $z \ll 1$, calculated from model fits to photohadronic cross
section data \cite{psb76,ss99}, is plotted in Figure 1.

UHECR Fe is seen to have a comparable GZK-type cutoff, but here at
$\approx 2\times 10^{20}$ eV, with photopair losses playing a minor
role. 

A realistic calculation of UHECR ion spectra in the evolving
background radiation fields must follow a reaction pathway
using Monte Carlo techniques, with the actual effective energy-loss
pathlength taking into account the EBL intensity
contributed by stars and black holes, then sometimes 
reprocessed through dust and gas. The exact form of the EBL intensity
between $\approx 1$ and 100 $\mu$ is poorly known, but can be
constrained by empirical galaxy SEDs and $\gamma$-ray observations.

\begin{figure}[t]
\center
\includegraphics[scale=0.35]{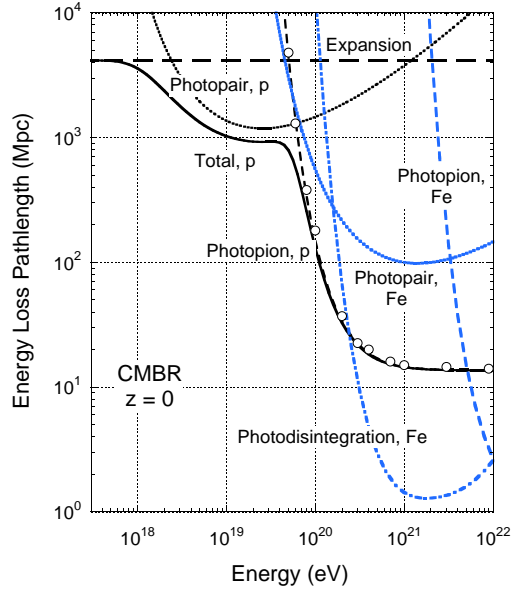}
\caption{ Energy-loss mean-free path for UHECR
protons (black) and Fe ions (gray) in the CMBR at $z = 0$, as a
function of total particle energy. The short-dashed black curve give
the analytical approximation for the UHECR proton photopion
energy-loss pathlength, eq.\ (\ref{eq1}), in comparison with the
numerical results, shown by open circles \cite{sta00}.  Numerical
results for the photopair energy-loss pathlength of UHECR protons and
Fe are shown by the dotted curves.  The gray dot-dashed curve gives
the photodisintegration energy-loss curve of Fe. Expansion losses are
shown by the long-dashed line, with $c/H_0 = 4170$ Mpc. }
\label{fig01}
\end{figure}


From the curves shown in Figure 1, we see that for either UHECR
protons or Fe ions, the sources of $E \gtrsim 10^{20}$ eV cosmic rays
must be found within a few hundred Mpc, and the $3\times 10^{20}$ eV
cosmic ray detected by the Fly's Eye \cite{bir95} must have
originated from a source within a few tens of Mpc. Both gamma-ray
bursts (GRBs) and radio-loud active galactic nuclei
(AGNs\footnote{Blazars are radio-loud AGNs viewed along the jet axis;
here the two terms are used interchangeably.})  satisfy this
requirement. Besides the availability of extragalactic sources within
tens of Mpc, several other conditions must also be met, including
\begin{enumerate}
\item sufficient power to make the UHECRs at the observed intensity; 
\item a plausible mechanism to accelerate particles to 
$\gtrsim 10^{20}$ eV energies; and
\item a model that can reproduce the observed UHECR spectrum,
 consistent with astronomical properties of the sources.
\end{enumerate}

The Auger clustering observations \cite{aug07} suggest new
calculations to determine the intergalactic medium (IGM) magnetic
field.  If the 2 (out of 27 total) UHECRs with arrival directions 
within 3$^\circ$ of Cen A are hypothesized to originate from 
 Cen A, then lower values of the mean IGM field  can be derived, 
$\gtrsim 10^{-10}$ G,
between us and 
Cen A, and $\gtrsim 10^{-11}$ G for sources at 75 Mpc. 
 The clustering observations and
the GZK length are tightly coupled concepts \cite{sta95}, 
as is established with the 
Auger data. Here we give new proton and ion energy loss calculations
after revisiting the problem of the EBL.

The Auger discovery \cite{aug07} that the 
arrival directions of $>56$ EeV are correlated with nearby, $\lesssim 75 $
Mpc AGNs, in particular, 2 cosmic rays within $3^\circ$ of Centaurus
A, does not mean that AGNs
are the sources of UHECRs, any more than O and B stars correlated with
galactic $\gamma$-ray sources mean that cosmic rays are accelerated by
high-mass stars \cite{tor03}. Long duration GRBs trace sites of active
star formation, and may be correlated to AGN
activity. AGNs trace the local matter distribution due to structure
formation, as do radio galaxies and GRBs, or for that matter clusters
of galaxies and radio-quiet Sy AGN.  Here we make the case for GRB and
radio-loud AGN/blazars as the sources of the UHECRs.  
UHECR ion acceleration in clusters of galaxies could also make a contribution
at $\lesssim 10^{19}$ eV, or from dim quasars \cite{bg99}, 
but this analysis forces us to reject UHECR origin from radio-quiet AGNs.

Effects of UHECR hadron acceleration can be varied,
including anomalous $\gamma$-ray emissions and characteristic
behaviors of GRB X-ray light curves.  Detection of high-energy
neutrinos from hadronic interactions at the sources will provide the
most definitive evidence for the presence of UHECRs \cite{hal07}.
Detection of cosmogenic GZK neutrinos from the decay of pions formed
in GZK interactions with CMBR photons gives a calorimetric measure of
UHECR power throughout cosmic time, with a characteristic spectrum
imprinted by production and propagation effects. ANITA and successor
Askaryan-effect detectors should soon be able to constrain a
long-duration GRB model of UHECRs, if they were protons.  Metals in
the composition of UHECRs change the spectral predictions, as well as
the GZK neutrino predictions \cite{ave04,anc07a}. The result of UHECR
ions vs. protons is mainly to enhance neutron $\beta$-decay neutrinos
but not significantly change the $\sim$ EeV neutrino flux.

Joint analysis of UHECR spectral, composition, and directional
information, now given most accurately with the hybrid Auger, with
neutrino and $\gamma$-ray data from GLAST, AGILE, and the ground-based
air and water Cherenkov detectors, should lead to a definitive
solution to the
problem of UHECR origin. A black hole origin is explored here.
Crucial to progress are the advances in
detector capabilities and instrumentation, multiwavelength
observations and multi-disciplinary analyses.

\section{Extragalactic Origin and Source Energetics}

The Larmor radius of an UHECR ion \cite{hil84} is 
$$r_{\rm L} = {E\over QB} \cong 1.1\;{(E/10^{19}{\rm~eV})
\over (Z/10) B(\mu {\rm G}) }\;{\rm kpc}$$
\begin{equation}
\cong 600\;{(E/6\times 10^{19}{\rm~eV})
\over (Z/10)\;B_{-11}}\;{\rm Mpc}\;,
\label{rL}
\end{equation}
where $B = 10^{-11}B_{-11}$ G is the mean magnetic field in which the
ion propagates. A significant anisotropy in the arrival directions of
UHECRs with energies exceeding $\approx 10^{19}$ eV might be expected
from Galactic sources along the Galactic plane or towards the Galactic
Center, even if they were composed primarily of Fe ($Z
= 26$), because any likely source class (pulsars, supernovae,
microquasars, high mass stars) would be confined to the thick disk of
the Galaxy. HiRes found no evidence for small-scale or large-scale
clustering \cite{abb05}. The Auger discovery \cite{aug07} of 
UHECR clustering along the supergalactic plane, anticipated
in analysis of AGASA, Haverah Park, and Yakutsk data \cite{sta95},  
 immediately rules out a galactic origin of the UHECRs. 

We now calculate
the emissivity $\dot{\cal E}$ required to power UHECRs,
assuming (incorrectly) that
UHECRs are protons that suffer expansion cooling and photopion and
photopair losses on the CMBR.  
The Auger observations \cite{yam07} of the spectrum of UHECRs, written
in the form $J_{24} = E^3 J/(10^{24}$ eV$^2$ m$^{-2}$ s$^{-1}$
sr$^{-1}$), where $J$ is the UHECR number intensity, implies that the
energy density of UHECRs with energy $E_{20}$ is $ u_{uhecr} \cong
6.7\times 10^{-22}J_{24}/E_{20} $ ergs cm$^{-3}$.  Using the energy-loss pathlengths shown
in Fig.\ 1 gives\footnote{This is the {\it proton} GZK radius.  Ion 
energy-loss pathlengths are calculated later in the paper.} $\dot {\cal
E} = u_{uhecr}/t_{tot} = cu_{uhecr}/r_{tot}$, so
\begin{equation}
\dot {\cal E} \;\big({{\rm ergs}\over {\rm Mpc}^{3}\mbox{-}{\rm~yr}}\big)\;=\;
 {6.0\times 10^{45}\over r_{tot}({\rm Mpc})}\;{J_{24}\over E_{20}}\;
\;.
\label{dotcalE}
\end{equation}
The  
sources of UHECRs with energy  $\approx 10^{20}E_{20}$ eV
must provide an emissivity $\dot{\cal E}_{44}$, 
in units of $10^{44}$ ergs Mpc$^{-3}$ yr$^{-1}$,
 given by the values shown in Table 1. 

 \begin{table}
\begin{center}
\caption{UHECR Source Emissivity}
\begin{tabular}{cccc}
\hline
$E$ (eV) & $J^a_{24}$ & $r_{tot}$(Mpc) & $\dot{\cal E}_{44}$ \\ 
\hline
\hline
$10^{20}$ & 0.6 & 140 & 0.4  \\
$10^{19.5}$& 1.8 & 900 & 0.4  \\
$10^{19}$ & 1.4 & 1000 &  0.8 \\
$10^{18.5}$ & 1.0 & 1700 & 1.2 \\
$10^{18}$ & 2.0 & $\cong 4000$ & 3.0  \\
\hline
\end{tabular}\par
\noindent $^a$From Auger data \cite{yam07}.
\end{center}
\end{table}

The required emissivity, from Table 1 using the Auger intensity, is
$\approx few \times 10^{44}$ ergs Mpc$^{-3}$ yr$^{-1}$ for $E\gtrsim
10^{19}$ eV, and $\approx 10^{45}$ ergs Mpc$^{-3}$ yr$^{-1}$ for
$E\gtrsim 10^{18}$ eV. Between $10^{18}$ eV and $10^{20}$ eV, the
increased energy density in UHECRs at the lower energies of this range
is balanced by the larger energy-loss length, thereby making the
injection emissivity roughly constant with energy.  At $E \lesssim
10^{18}$ eV, the emissivity increases roughly proportional to energy
because the energy-loss pathlength is approximately equal to the
Hubble radius at $10^{18}$ eV, and $E^2 J \propto 1/E$ at these
energies.

Classical long duration GRBs have a volume- and time-averaged
emissivity in the X-ray/soft $\gamma$ ray (``$X/\gamma $") band
comparable to this value, a coincidence first noted by Vietri
\cite{vie95} and Waxman \cite{wax95}. We can reproduce this estimate
by noting that the average $\gtrsim 20$ keV fluence of BATSE (the
Burst and Transient Source Experiment (BATSE) on the {\it Compton
Observatory}) GRBs is $\approx 10^{-5}$ ergs cm$^{-2}$ \cite{pre00},
and that there are about 500 long-duration BATSE GRBs over the full
sky per year \cite{ban02}, giving an average GRB flux of $\approx
10^{-2}$ ergs cm$^{-2}$ yr$^{-1}$.\footnote{Gonz\'alez (private
communication, 2003), calculates a total fluence per year of
$0.63\times 10^{-2}$ ergs cm$^{-2}$ for 1293 GRBs in the 4th BATSE
catalog, including GRBs of both long- and short-duration, considering
666 GRBs/yr full sky, implying a bolometric average energy density
$\lesssim 10^{-20}$ ergs cm$^{-3}$ of GRB light.}  BATSE GRBs are, on
average, at redshift $z \approx 1$ (Hubble radius $R_{\rm H} \cong
4200$ Mpc), so that the $\gtrsim 20$ keV emissivity of long duration
GRBs is $$\dot{\cal E}_{GRB} \approx {4\pi R_{\rm H}^2 \times
10^{-2}{\rm ~ergs~cm}^{-2}{\rm~yr}^{-1}
\over 4\pi R_{\rm H}^3/3}\cong 
$$
\begin{equation}
7\times 10^{43}{\rm ~ergs~Mpc}^{-3}{\rm~yr}^{-1}\;,
\label{dotEGRB}
\end{equation}
in rough agreement with the UHECR emissivity requirements shown in 
Table 1. 

Two assumptions (at least) underlay this estimate: 
One is that an average emissivity can apply to the peculiar 
emissivity of the local 
$\lesssim 100$ Mpc sphere that we inhabit. 
Greater GRB activity at $z\gtrsim 1$
compared to the present epoch means that there would be fewer sources
within the GZK radius \cite{ste00}.  On the other hand, additional
classes of GRB sources, in particular, the X-ray flashes or the low
luminosity GRBs \cite{lia07}, can provide a substantial addition to
the emissivity. Wang et al.\ (2007) \cite{wan07} estimate a local
 emissivity $\dot{\cal E}
\approx 250\times 10^{44}$ ergs Mpc$^{-3}$ yr$^{-1}$
 in the kinetic output of nearby low-luminosity GRB hypernovae (which 
exceeds the emisivity in observed $\gamma$ rays). A further assumption is that there is
comparable energy injected in UHECRs as detected in electromagnetic
radiation. This could well be wrong. A large, $\approx 30$ -- 100,
 baryon loading is
required if long duration BATSE/Beppo-SAX/GBM--type GRBs 
are the sources of UHECRs \cite{wda04}.

\begin{figure}[t]
\center
\includegraphics[scale=0.48]{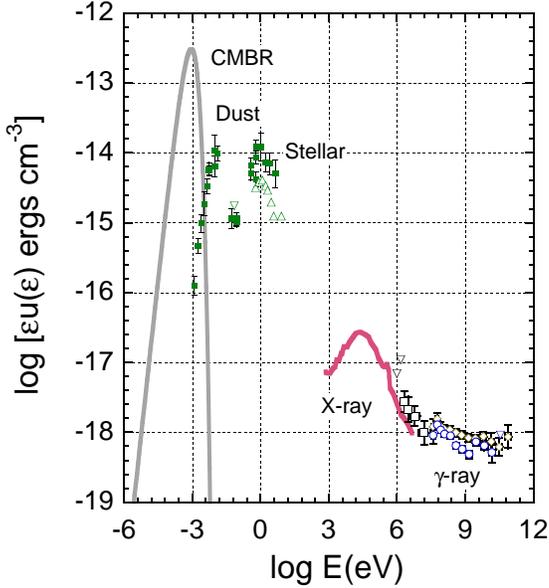}
\caption{Intensity of extragalactic background light from 
the microwave through the $\gamma$-rays, with components as 
labeled.}
\label{fig02}
\end{figure}

A similar emissivity estimate for blazar AGNs can be made on the basis
of the unresolved/diffuse extragalactic $\gamma$ radiation (DEGR)
observed with the Energetic Gamma-Ray Experiment Telescope (EGRET) on
the {\it Compton Observatory} \cite{sre98,smr04} (see Fig.\
\ref{fig02}). Between $\approx 100$ MeV and $\approx 100$ GeV, the
``$\nu F_\nu$" intensity $\e I_\e \approx 1.5$ keV cm$^{-2}$ s$^{-1}$
sr$^{-1}$, so that the diffuse $\gamma$-ray energy flux received here
at Earth is $\approx 0.5$ erg cm$^{-2}$ yr$^{-1}$.  Because blazar AGNs
comprise the largest number of identified EGRET sources, they
undoubtedly make up a large fraction of this radiation, with estimates
ranging from $\approx 20$\% to nearly 100\%.

It is important here to distinguish between the nearly lineless BL Lac
objects and the flat spectrum radio quasars (FSRQs) with strong broad,
optical emission lines. The BL Lac objects detected at $\gamma$-ray
energies are at mean redshift $\langle z \rangle \approx 0.3$, whereas
the FSRQs detected with EGRET are at $\langle z \rangle \approx
1$. The BL Lac and FSRQ contributions to the DEGR are found
\cite{der07} from model fits to the redshift and peak flux
$\gamma$-ray data to contribute at the $\sim 5$\% and $\sim 15$\% of
the total diffuse $\gamma$-ray intensity, respectively.

Following the reasoning leading to eq.\ \ref{dotEGRB}, the mean
emissivity in $\gamma$ rays is $\dot{\cal E}_{BL~Lac} \approx
10^{45}{\rm ~ergs~Mpc}^{-3}{\rm~yr}^{-1}$ for BL Lac objects, and
$\dot{\cal E}_{FSRQ} \approx 6\times 10^{44}{\rm
~ergs~Mpc}^{-3}{\rm~yr}^{-1}$ for FSRQs.  On the basis of energetics
arguments, both FSRQs and BL Lacs appear to have sufficient energy to
power the UHECRs. The more powerful FSRQs are, however,  rare in our
local vicinity, so that average emissivity becomes a mistaken 
concept. The nearest high luminosity FR II radio galaxies
associated with FSRQs are at $z\approx 0.1$; the FR II radio galaxy
Cygnus A has $z \approx 0.056$, so is $\approx 250$ Mpc distant, and
is far outside the GZK radius for a $10^{20}$ eV proton.

By comparison, many low-luminosity FR I radio galaxies associated with 
BL Lac objects are found nearby. For example, Centaurus A and M87 
are $\approx  4$ and $\approx 16$ Mpc distant, respectively. These 
FR Is have lower jet power than the powerful FR IIs. This makes it 
more difficult to accelerate protons to ultra-high energies, though 
not heavy ions, as we now show. 

\section{Acceleration to Ultra-High Energies}

Acceleration in relativistic blast waves can proceed through a number
of mechanisms. First-order shock Fermi acceleration, analogous to the
standard particle acceleration mechanism in supernova remnant shocks,
is the obvious process \cite{kd99}. But acceleration of particles to ultra-high
energies through first-order processes at a relativistic external
shock encounters kinematic difficulties to reach ultra-high energies
\cite{ga99,gak99}.  When particles diffuse upstream from the shock,
the shock overtakes the particles before they can change their
direction by more than an angle $\approx 1/\Gamma$, where $\Gamma$ is
the Lorentz factor of the blast wave. This prevents the particle from
increasing its energy by more than a factor $\approx 2$ in each cycle
following the first particle capture by the blast wave.  Even so,
acceleration to UHE through first-order relativistic shock
acceleration can take place if the surrounding medium is strongly
magnetized, as could be the case in the winds of a massive Wolf-Rayet
progenitor.  If the colliding shells of blazars and GRBs are
sufficiently magnetized, first-order processes can also apparently
accelerate protons to UHEs \cite{wax95}.

Here we present two arguments showing that UHEs can be achieved in the
relativistic shocks of GRBs and blazars. The first considers
stochastic acceleration in the external shock of a GRB or blazar, and
the second considers general source power requirements \cite{wax04}
for particle acceleration in GRBs, blazars, or other candidate UHECR
sources.

\subsection{A. Stochastic Acceleration in the Blast Wave Shell}

Second-order processes in an external shock of a blast wave can be
shown to permit hadronic acceleration to ultra-high energies, in 
comparison with radiative and expansion losses and escape \cite{dh01}.
Using parameters appropriate to a GRB, consider a blast wave with
apparent isotropic energy release $E_0 = 10^{54}E_{54}$ ergs,
(initial) coasting Lorentz factor $\Gamma_0 = 300\Gamma_{300}$, and
external medium density $n_0 = 100n_2$ cm$^{-3}$.  The comoving
blast-wave volume for a spherically symmetric explosion at distance
$x$ from the center of the explosion is
\begin{equation}
V^\prime = 4\pi x^2 \Delta^\prime,
\end{equation}
where primes refer to the comoving frame, and the shell width
$\Delta^\prime = x/12\Gamma$ (the factor $1/12 \Gamma$ is the product
of the geometrical factor $1/3$ and the factor $1/4\Gamma$ from the
compression of the material by the shock, in accord with the
conservation laws of relativistic hydrodynamics).

A necessary condition to accelerate to some energy $E^\prime_{max}$ is
that the particle Larmor radius is less than the size scale of the
system \cite{hil84}, that is, 
$r^\prime_{\rm L} \lesssim {x_d/ 12\Gamma}\;.$
$E_{max}$ in the stationary explosion frame is
then given by
\begin{equation}
r_L^\prime = {E_{max}^\prime\over ZeB^\prime } =
 {E_{max}\over \Gamma Ze B^\prime } < \Delta^\prime.
\end{equation}
The largest particle energy is reached at the deceleration radius $x =
x_d$ when $\Gamma \cong \Gamma_0$, where the deceleration radius
\cite{mr92} $$x_d \equiv ( {3 E_0\over 4\pi\Gamma_0^2 m_p n_0})^{1/3}
\cong$$
\begin{equation}
 2.6\times 10^{16} ({E_{54}\over \Gamma_{300}^2 n_2})^{1/3}\;\rm{cm}\;.
\end{equation}
Hence $
E_{max} \cong {Z eB^\prime x_d/ 12}\;.
$

The mean magnetic field $B^\prime$ in the GRB blast wave is assigned
in terms of a magnetic field parameter $\e_B$ that gives the magnetic
field energy density in terms of the energy density of the downstream
shocked fluid, so $$B^\prime = (32\pi n_0 \e_B
m_pc^2)^{1/2}\sqrt{\Gamma(\Gamma-1)}\cong $$
\begin{equation}
 0.4 \sqrt{\e_B n_0} \;\Gamma
 \;\cong\; 1200 \sqrt{\e_B n_2}\,\Gamma_{300}\;{\rm G}\;.
\end{equation}
Thus 
\begin{equation}
E_{max} \cong 8\times 10^{20} Z n_2^{1/6} \e_B^{1/2} 
\Gamma_{300}^{1/3} E_{54}^{1/3}\;{\rm eV}
\label{Emax}
\end{equation}
\cite{vie98,dh01}, 
so that external shocks of GRBs can accelerate particles 
to ultra-high and, indeed, super-GZK energies. The highest energies
are more easily reached for ions of large $Z$, which helps to relax
limits on density and $\e_B$.

For values appropriate to blazar AGNs, 
acceleration to  UHEs in the external shock of a blazar plasma jet
seems possible from eq.\ \ref{Emax}, at
least for FSRQs. For this blazar class, superluminal motion
observations \cite{vc94} and constraints from the requirement that the
emission region be optically thin to $\gamma\gamma$ pair production
attenuation \cite{mgc92} imply $\Gamma \sim 10$ -- 30.  The brightest
blazar flares observed with EGRET reach $\approx 10^{49}$ ergs
s$^{-1}$ and last for $\lesssim 1$ day \cite{muk97}, suggesting again
that the total isotropic energy release $E_{54}\approx 1$. For $\Gamma
\approx 30$, the deceleration time $t_d = (1+z)x_d/\Gamma^2 c$ is
shorter than a day, provided that the blazar blast wave passes through
a medium of density $\approx 10^2$ cm$^{-3}$.  With these
numbers, UHECRs can be accelerated in the external shocks of FSRQs.
Much improved data for this estimate will be provided with {\it GLAST}.

The case for UHECR acceleration in BL Lac objects is not as favorable.
Although spectral modeling suggests that $\Gamma$ could be as large as
50 \cite{kca02}, values of $\Gamma$ inferred from superluminal motion
observations rarely exceed 5 -- 10 \cite{jor05}. 
The mean distances, energy
fluxes, and flare durations are generally smaller for BL Lacs than
quasars, and the total energy in
BL Lac $\gamma$-ray flares is smaller than for FSRQ flares. 
The deconvolved apparent $\gamma$-ray luminosity from PKS 2155-304 
 flares during the 2006 July flaring state was
 $\approx 3\times 10^{45}$ erg s$^{-1}$ \cite{aha07a}, and lasts between $\approx 10^2$ and 
$10^3$ s, giving a total apparent energy releases $\approx 3\times 10^{47}$ -- $3\times 10^{48}$
ergs.
The surrounding medium density is lower in BL Lacs
than in FSRQs, given that blazars have smaller optical emission-line
equivalent widths than FSRQs, and therefore probably a smaller column
density of broad line region material. All these factors 
make it harder for
low-$Z$ ions to reach UHEs. So, if FR I radio galaxies are the sources of 
UHECRs, then they would have a suppressed proton and low-$Z$
ion content due to acceleration limitations.

Why, however, consider stochastic acceleration processes in the blast waves
formed by the external shocks in blazars and GRBs, given that
first-order Fermi acceleration through colliding shells, which can
operate in both blazars and GRBs, could accelerate the cosmic rays to
high energies?  The problem here is that the material which forms the
outflowing plasma winds is processed through extreme environments
around supermassive black holes of blazar AGNs and newly formed black
holes in GRBs. Whether ions can survive, as required by the Auger data
\cite{ung07}, is an open question.  If they do survive, then Wang,
Razzaque, \& M\'esz\'aros \cite{wrm07} have shown that
acceleration of UHECR ions to $\gtrsim 10^{20}$ without significant
losses to photodisintegration in GRB internal shocks, external shocks,
and hypernovae is possible (see also \cite{anc07a}).

In a colliding shell scenario, the ejecta originates from the vicinity
of a black hole. For the fireballs that power a GRB, heavier nuclei
are broken down into protons and neutrons, deuterium and $\alpha$
particles \cite{bel03}.  Nuclear breakdown reactions, either through
direct particle spallation processes or through photodisintegration
from the intense radiation field in the vicinity of a black hole,
make it questionable if baryonic material ejected from the central
engines of black holes in GRBs or blazars has any metal ($A > 4$) content.

In a colliding shell scenario, successive waves of ejecta are
supposed to form the flares in blazars and GRBs, and if the hadronic
content in the ejecta were composed entirely of protons and light
nuclei, then such an acceleration scenario could not account for the
Auger results.  Wang, et al.\ \cite{wrm07}
argue, however, that mixing instabilities could develop to seed the
relativistic ejecta with heavier ions as the blast wave passes through
the the stellar envelope surrounding a GRB.  In this case, the amount of baryon
contaminant would have to be carefully regulated without heavily
loading and quenching the fireballs, unless dirty
fireball bursts \cite{dcb99} and quenched, or choked \cite{mw01}
fireballs also occur.

A scenario involving an external shock would permit the capture of
ions from the surrounding medium. This medium could be highly enriched
from the presence of circumnuclear starbursts surrounding the
supermassive black hole in a blazar AGN or OB associations in which a
GRB might reside, and so have a considerable ion content.  Even if the material had
sub-Solar metallicity, heavier ions would be preferentially
accelerated due to their larger charge $Z$ (see eq.\ \ref{Emax}),
 producing a mixed composition in the UHECRs.

\subsection{B. Power Requirements for Electrodynamic Acceleration}

We apply Waxman's argument \cite{wax04}, giving minimum source
power to accelerate $10^{20}E_{20}$ eV protons, to ions. 
In a region of size
$R$ and magnetic field $B$, electromagnetic forces can accelerate a
particle to a maximum energy of $E_{max} > E_{par} \simeq Ze\beta
BR$. The available time in the comoving frame is shortened by bulk
Lorentz factor $\Gamma$, so that the effective size for acceleration
is $\approx R/\Gamma$, and $BR>\Gamma E_{par}/Ze\beta$. The required
power of the magnetized flow is 
$$L \approx 2\times 4\pi R^2 v\times {B^2\over 8\pi }
\approx \beta c (BR)^2 \approx{c\Gamma^2 E_{par}^2\over Z^2 e^2 \beta}$$
or 
$$L\cong{3\times 10^{45}\over Z^2}\;{\Gamma^2\over \beta}
\;E_{20}^2\;{\rm~ergs~s}^{-1}\;$$
\begin{equation}
\cong {5\times 10^{42}
\over (Z/26)^2}\;{\Gamma^2\over \beta}
\;E_{20}^2\;{\rm~ergs~s}^{-1}\;,
\label{L}
\end{equation}
including a factor of 2 for the plasma jet kinetic
power.
If the nonthermal luminosity is a good measure of jet power, then
we can decide whether different source classes are good candidate
UHECR sources.

For GRBs, $\G \approx 300$, and $L_{GRB}\gg
10^{50}(\G/300)^2E_{20}^2/Z^2$ ergs s$^{-1}$.  Apparent isotropic
$X/\gamma$ \ powers of long-duration GRBs are regularly measured in
excess of $10^{50}$ ergs s$^{-1}$ \cite{fb05}, so long-duration GRBs are a viable
candidate for UHECR acceleration. For the low-luminosity GRBs, which
may only reach $\approx 10^{48}$ -- $10^{50}$ ergs s$^{-1}$ \cite{sod04}, higher
$Z$ ions can still be accelerated to super-GZK energies if $\Gamma$
remains large. It will be interesting to compare radiative powers with
this minimum power using allowed values of $\Gamma$ determined from
GLAST data through $\gamma\gamma$ opacity arguments, for different
classes of GRBs.

For blazars with $\Gamma \cong 10$, $L_{blazars}\gg 3\times
10^{47}(\G/10)^2 E_{20}^2/Z^2$ ergs s$^{-1}$.  FSRQ blazar
$\gamma$-ray flares brighter than $\simeq 10^{47}$ ergs s$^{-1}$ were
frequently detected with EGRET \cite{har99}.  Present data does not
exclude FSRQ blazars from being the sources of UHECRs, especially if
the accelerated UHECRs are primarily heavy ions. Comparisons \ between
measured $\gamma$-ray luminosity and minimum power requirements using
values of $\Gamma$ obtained from $\gamma\gamma$ arguments, both of
which can be determined from GLAST LAT data, can decide whether FSRQ
blazars can accelerate UHECR protons and ions.

Eq.\ (\ref{L}) shows that it is more difficult to accelerate UHECRs,
especially UHECR protons, in the lower luminosity X-ray/TeV blazars
with apparent $\gamma$-ray powers $\lesssim 10^{45}$ ergs s$^{-1}$.
If TeV observations with VERITAS, HESS, or MAGIC were to require,
either from spectral modeling or $\gamma\gamma$ arguments, that
$\Gamma \gtrsim 50$ in sources like Mrk 421, Mrk 501, or PKS 2155-304,
acceleration even of Fe to super-GZK eneriges 
might be problematic in the BL Lac sources.

It is also interesting to apply eq.\ (\ref{L}) to parameters 
of merging clusters of galaxies, which have also been studied as a
potential source of UHECRs \cite{bbp97,ino07}.

The gain in kinetic energy when the minor cluster of mass $M_2= 10^{14}M_{14} M_\odot $,
treated as a test particle in the mass distribution of the dominant
cluster of mass $M_1 = 10^{15}M_{15} M_\odot $, falls from radius
$r_1$ to radius $r_2 (\leq r_1) $, is $M_2 v_2^2/2 = GM_1 M_2
(r_2^{-1} -r_1^{-1}$) (e.g., \cite{bd05}).  Thus $v\cong
\sqrt{2GM_1/r_2}$ when $r_2 \ll r_1$, and
\begin{equation}
v_2 = \beta_{cl} c \approx 3000 \;\sqrt{M_{15}\over r_2({\rm Mpc})}\kms,
\label{vsim}
\end{equation}
so $\beta_{cl} \simeq 10^{-3}$.  If $r_2$ is scaled to a typical core
radius of the dominant cluster, $\approx 0.25$ Mpc, the greater power
output during this merging episode occurs during a timescale shorter
by a factor $r^{3/2}$.  This can be shown by noting that the
characteristic merger time $\hat t$ is determined by the acceleration
$a = GM_1/r_1^2$ at the outer radius.  Because $ r_1 \approx a_2 \hat
t^2/2$,
\begin{equation}
\hat t \cong \sqrt{{2r_1^3\over GM_1}}\approx 660\; {r^{3/2}_{\rm Mpc}
 \over M_{15}}\;{\rm~Myr}\;,
\label{tcongr}
\end{equation}
where $r_1 = r_{\rm Mpc}$ Mpc.  The available
energy in the collision is
\begin{equation}
{\cal E} \approx {GM_1 M_2\over r_1} 
\approx {8\times 10^{63}\over r_{\rm Mpc}}\; M_{15}\; 
M_{14}\;\ergs\;.
\label{calE}
\end{equation}

The ratio of eqs.\ (\ref{calE}) and (\ref{tcongr}) gives the maximum
power available from the gravitational energy of the merging clusters,
namely
\begin{equation}
L_{cl} = {{\cal E}\over \hat t} \cong 4\times 10^{47}\; 
{M_{15}^2 \over r^{5/2}_{\rm Mpc}}\;M_{14}
~\ergss,
\label{Lcl}
\end{equation}
When $r_{\rm Mpc}\sim 0.25$,
corresponding to typical core radii of galaxy clusters like Coma,
the maximum merger power is $\approx 30$ larger and 
the timescale, eq.\ (\ref{tcongr}), is a factor $\approx 8$ less.  The
luminosity requirement to accelerate UHECRs with energy $E_{20}$ is,
for parameters typical of merging cluster, $$L_{gc} \gtrsim 3\times
10^{48} E_{20}^2/[Z^2 (\beta_{cl}/10^{-3})]\ergss\;,$$ from eq.\
(\ref{L}). 
According to this criterion, it is not difficult to
accelerate UHECR protons in galaxy cluster environments. 
This estimate does not consider actual timescales \cite{ino07a} of 
acceleration, which result in maximum proton 
energies $\lesssim 10^{19}$ eV for p from 
nonrelativistic shocks \cite{bd03}.  
 Acceleration of p to $\lesssim 10^{19}$ eV 
and Fe to $\lesssim 10^{20}$ eV is possible \cite{ino07} 
in large Mach number
\cite{gb03} cluster accretion shocks.

The time for the merger during maximum power output, from eq.\
(\ref{tcongr}), corresponds to a distance $\hat r = c \hat t \cong 25
(r_{\rm Mpc}/0.25{\rm~ Mpc}  )^{3/2}/M_{15}$ Mpc, and 
UHECR Fe might go through some significant photo-erosion in the CMBR and
EBL,  so Fe would have difficulty surviving
to $\gtrsim 10^{20}$ eV if it were accelerated by merger shocks
in merging clusters of galaxies (cf. \cite{ino07a} 
for cluster accretion shocks).

\section{X-ray and $\gamma$-ray Signatures of UHECR Acceleration}

Indirect evidence for UHECR acceleration is given by analysis of
spectra and light curves of GRBs and blazars.  In the relativistic
blastwave framework, the hard $X/\gamma$ radiation during the prompt
phase of a GRB is primarily nonthermal synchrotron radiation,
including perhaps some thermal photospheric emission \cite{pee07}. A synchrotron
self-Compton (SSC) component would accompany the nonthermal
synchrotron emission, and display a correlated behavior.

The delayed hard emission tail in GRB 940217 \cite{hur94}, lasting for
over 90 minutes after the start of the GRB, could be a signature of a
long lasting, hadronic acceleration process.  The slower decay of the
hadronic emission component compared to the leptonic component, as
expected from the standard blast wave model \cite{bd98}, might explain
the delayed emission. Nevertheless, a leptonic model with an SSC
component appearing in the GeV regime when the synchrotron
component has decayed to optical/UV energies could also account for
the delayed behavior of GRB 940217, or the superbowl burst, GRB 930131
\cite{som94}.

A more difficult case to explain in terms of nonthermal lepton
radiations is GRB 941017, observed with the BATSE Large Area Detector
(LAD) and the EGRET Total Absorption Shower Counter (TASC). Gonz\'alez
et al.\ \cite{gon03} reported the detection of an anomalous MeV emission
component in the spectrum of this burst that decays more slowly than
the prompt emission detected with the LAD between $\approx 50$ keV and
1 MeV range. The multi-MeV component lasts for $\gtrsim 200$ seconds,
and is detected both with the BATSE LAD near 1 MeV and with the EGRET
TASC between $\approx 1$ and 200 MeV. The spectrum is very hard, with
a photon number flux $\phi(\epsilon_\gamma)\propto
\epsilon_\gamma^{-1}$, where $\epsilon_\gamma $ is the observed photon
energy.

This component was not predicted nor is easily explained within the
standard leptonic model for GRB blast waves. It has been suggested
that this emission component could
be related to Comptonization of reverse-shock photons by the forward
shock electrons, including self-absorbed reverse-shock
optical synchrotron radiation \cite{pw04}. Extremely large apparent isotropic
energies are however required. 

This component could be a consequence of the acceleration of
ultrarelativistic hadrons at the relativistic shocks of GRBs
\cite{da04}. A pair-photon cascade initiated by photohadronic
processes between high-energy hadrons accelerated in the GRB blast
wave and the internal synchrotron radiation field produces an emission
component that appears during the prompt phase, as shown in Fig.\
3, but delayed due to the time required for acceleration. 
Photomeson interactions in the relativistic blast wave would simultaneously
make a beam of UHE neutrons and neutrinos, as proposed for blazar
jets \cite{ad03}. Subsequent photopion production of these neutrons
with photons outside the blast wave will produce a directed
hyper-relativistic electron-positron beam in the process of charged
pion decay and the conversion of high-energy photons formed in $\pi^0$
decay. These energetic leptons produce a synchrotron spectrum in the
radiation reaction-limited regime extending to $\gtrsim$ GeV energies,
with properties in the 1 -- 200 MeV range similar to that measured
from GRB 941017. GRBs displaying anomalous $\gamma$-ray components are
most likely to be detected as sources of high-energy neutrinos with
IceCube.

\begin{figure}[t]
\includegraphics[scale=0.39]{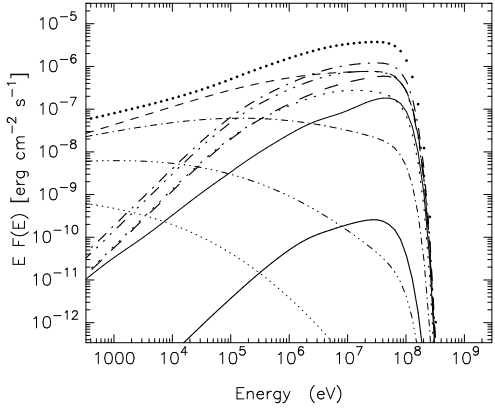}     
\caption{Atoyan's calculation of photon energy flux 
from an electromagnetic cascade initiated
by photopion secondaries in a model GRB, with parameters given in
Ref.\ \cite{da04}. Five generations of Compton (heavy curves) and
synchrotron (light curves) are shown. The first through fifth
generations are given by solid, dashed, dot-dashed,
dot-triple--dashed, and dotted curves, respectively. The total cascade
radiation spectrum is given by the upper bold dotted curve. }
\end{figure}

If UHECRs are accelerated by GRB blast waves, then blast-wave dynamics
will be affected by the loss of internal energy when the UHECRs
escape.  This effect \cite{der07a} could explain the rapid X-ray
declines in the Swift GRB light curves \cite{tag05}. Protons
undergoing photopion interactions with photons at the peak photon
frequency $\nu_{pk}$ or peak dimensionless energy $\e_{pk} =
h\nu_{pk}/m_ec^2 \cong 2\Gamma \ep_{pk}/(1+z)$ of the $\nu F_\nu$
spectrum have energy, as measured by an observer in the explosion
frame,
\begin{equation}
E_{pk} = m_pc^2 \gamma_{pk}\simeq {3\times 10^{16} (\Gamma/300)^2\over
(1+z) \e_{pk}}\;{\rm eV}.
\label{Epk}
\end{equation} 
The comoving time required
for a proton with energy $E_{pk}$ to lose
a significant fraction of its energy through photohadronic processes is
$$t^\prime_{\phi\pi}(E_{pk}) 
\simeq 
{m_ec^2 x^2 \Gamma^2 \ep_{pk}\over K_{\phi\pi}
\sigma_{\phi\pi}d_L^2f_{\e_{pk}}}
$$
\begin{equation}
\simeq 
2\times 10^6\; {x_{16}^2 (\Gamma/300) (1+z) 
\e_{pk}\over d_{28}^2 f_{-6}}\;{\rm s}\;,
\label{tprimephipi}
\end{equation} 
where $x = 10^{16} x_{16}$ cm and $f_{\e_{pk}} = 10^{-6}f_{-6}$ ergs
cm$^{-2}$ s$^{-1}$ is the $\nu F_\nu$ flux measured at $\e_{pk}$; the
relation between $E_{pk}$ and $\e_{pk}$ is given by eq.\ (\ref{Epk}).

The dependence of the terms $x(t)$, $f_{\e_{pk}}(t)$, $\Gamma(t)$, and
$\e_{pk}(t)$ on observer time in eq.\ (\ref{tprimephipi}) can be
analytically expressed for the external shock model in terms of the
GRB blast wave properties $E_0$, $\Gamma_0$, environmental parameters,
e.g., $n_0$, and microphysical blast wave parameters $\e_B$ and $\e_e$
\cite{der07a}.  This can also be done for other important timescales,
for example, the (available) comoving time $\tp_{ava}$ since the start
of the GRB explosion, the comoving acceleration time $\tp_{acc} =
\zeta_{acc} m_pc^2 \gp/eBc $, written as a factor $\zeta_{acc}\gg 1$
times the Larmor timescale \cite{rm98} ($\zeta_{acc} = 10$ in the
Figure 3), the escape timescale $\tp_{esc}$ in the Bohm diffusion
approximation, and the proton synchrotron energy loss timescale
$\tp_{syn}$.

\begin{figure}[t]
\includegraphics[width=2.8in]{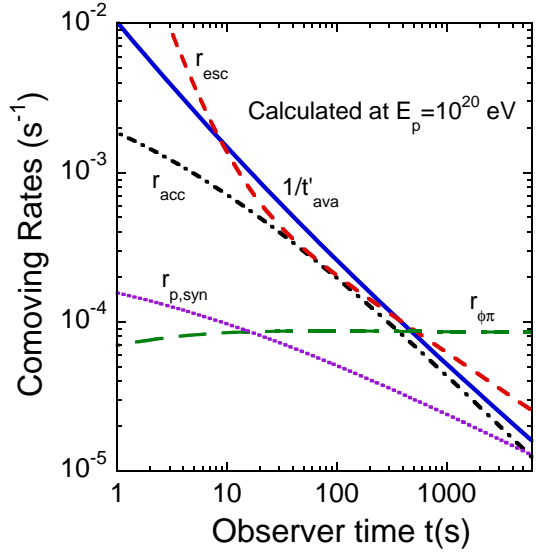}    
\caption{Rates and inverse timescales as a function of 
observer time for $10^{20}$ eV cosmic ray protons as measured 
by a stationary external observer. The figure uses
 parameters given in the text.}
\label{f4}
\end{figure}

Fig.\ \ref{f4} shows the rates (or the inverse of the timescales) for 
$10^{20}$ eV protons in the case of 
an adiabatic blast wave that decelerates in a uniform
surrounding medium. The parameters are
$$z = 1, \;\Gamma_{0} = 150,\; E_{54} = 10,\; n_0 = 1000{\rm~cm}^{-3}, $$
$$\;\e_{e} = 0.1, \;{\rm and~ } \e_{B} = 0.3\;.$$

For these parameters, it takes a few hundred seconds to accelerate
protons to energies $\approx 10^{20}$ eV, at which time photohadronic
losses and escape start to be important. Photohadronic losses inject
electrons and photons into the GRB blast wave. The electromagnetic
cascade emission, in addition to hyperrelativistic electron
synchrotron radiation from neutron escape followed by subsequent
photohadronic interactions \cite{da04}, makes a delayed anomalous
$\gamma$-ray emission component as observed in GRBs 940217 and 941017
\cite{hur94,gon03}. Ultra-high energy neutrino secondaries are
produced as by-products of the hadronically-induced cascade.  The
ultra-high energy neutrons and escaping protons, accompanied by escaping UHECR ions,
form the UHECRs with energies $\gtrsim 10^{20}$ eV.
Detection of high-energy neutrinos from GRBs would confirm the
importance of hadronic processes in GRB blast waves.

The GRB blast wave quickly loses internal energy due to the
photohadronic processes and particle escape. The blast wave will then
 decelerate, producing a rapidly decaying X-ray flux. I 
\cite{der07a} argue that the decaying fluxes in Swift GRBs are
signatures of UHECR acceleration by GRBs. (See \cite{zha06} 
for other explanations of the Swift data.) If this scenario is
correct, GLAST will detect anomalous $\gamma$-ray components
preferentially in those GRBs that undergo rapid X-ray declines in
their X-ray light curves. 

Anomalous $\gamma$-ray signatures have also been detected in the
spectra of blazar AGNs, for example, the orphan $\gamma$-ray flare
observed in the TeV blazar 1ES 1959+650 \cite{kra04}.  This is a case
where the correlated variability between the synchrotron X-rays and
SSC $\gamma$ rays expected in the standard synchrotron/SSC TeV blazar
model is not observed. Orphan X-ray flares from an hadronic emission
component could be produced by cosmic ray protons with Lorentz factors
$\gamma\approx 10^2$ -- $10^4$ undergoing photohadronic interactions
with reflected X-ray target photons
\cite{bot05}. The implied neutrino signature is unfortunately too weak
to be detected with IceCube \cite{brp05} or a Northern Hemisphere km-scale
neutrino telescope. 
In order for a TeV blazar to have efficient 
photohadronic interactions of UHECR protons with ambient jet synchrotron 
radiation,
the blazar must be optically thick to $\gamma\gamma$ pair production 
\cite{rmz04,drl07}, so we would only expect PeV neutrinos from 
TeV blazars during times of low-state TeV $\gamma$ ray flux. Important for 
this search is multiwavelength GRB and blazar capability.

In the case of FSRQ blazars, there is as yet no strong evidence for
anomalous $\gamma$-ray emission components that could be associated
with UHECR acceleration.  This lack of evidence should not be
considered definitive, for a number of reasons: (1) The leptonic
models for FSRQs are more complicated than for TeV blazars or GRBs,
and involve a variety of external radiation fields and, consequently,
more parameters. Back-scattered radiation in structured blazar AGN
environments makes another radiation feature. 
This makes it more difficult to ascribe an emission
component to a non-leptonic origin. (2) Except in a few
cases, e.g., \cite{weh98,har01}, the
$\gamma$-ray data from FSRQs taken during the EGRET era didn't have
extensive
multiwavelength monitoring.  (3) The sensitivity of the
EGRET telescope permitted spectral fits with a single power law integrated
over several days in all but a few cases. This situation will change
dramatically
with the launch of GLAST in 2008, and is already changing with 
the development of low-energy
threshold air-Cherenkov $\gamma$-ray detectors. A significant advance
in this direction was reported at this conference by the MAGIC
collaboration \cite{tes07}, namely the 6 and 5 $\sigma$ detections of
3C 279 ($z \cong 0.538$) in the respective bands $\sim 80$ -- 220 GeV
and $\gtrsim 220$ GeV. This opens the question of what other nearby
FSRQs will be detected with MAGIC, and what this means for the
intensity of the EBL.

\section{UHECR Protons from GRBs}

\begin{figure}
\includegraphics[width=2.6in]{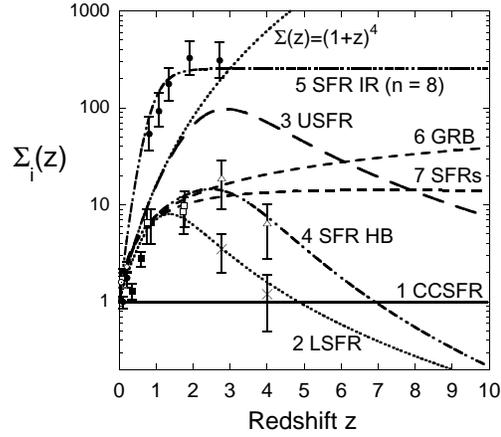}     
\caption{Different star formation rate (SFR) histories, as described in 
the text.}
\label{f5}
\end{figure}

\begin{figure}
\includegraphics[width=2.8in]{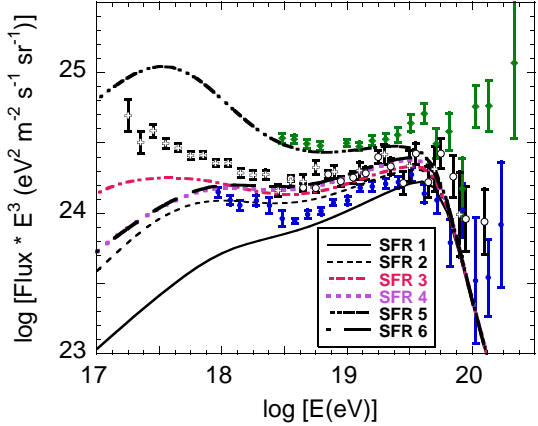}     
\caption{UHECR data from monocular HiRes  (open crosses and circles),
Auger (filled circles), and AGASA (filled diamonds) data, compared
with GRB predictions for UHECR protons with different SFRs shown in 
Fig.\ 5. Protons
are injected from $10^{14}$ eV to an exponential cutoff energy
$E_{max}=10^{20}$ eV with local emissivity $= 60\times 10^{44}$ ergs
Mpc$^{-3}$ yr$^{-1}$.}
\label{f6}
\end{figure}

GRBs are argued to be the sources of the UHECRs for over a decade
\cite{vie95,wax95,der02}.  Well-defined calculations based on particle
physics and GRB astronomy for UHECR proton propagation in the
expanding universe, treating expansion, and photopion and photopair
interactions with CMBR photons only (see Figure 1), give results that
can be compared with data and used to benchmark more detailed
calculations involving nuclei and various source classes.

Restricted to the long-duration class, of which we have the most
knowledge, the act of making a calculation of the UHECR spectrum from
GRBs requires (i) an injection spectrum, which we take to be a
power-law with number injection index $s = 2.2$ between $10^{14}$ eV
and an exponential cutoff energy, usually taken to be 
$10^{20}$ eV; (ii) a local emissivity, which acts as an overall
normalization factor on the UHECR spectrum; and (iii) a star formation
rate (SFR). Here the underlying assumption is that the rate-density of
GRBs follows star formation activity, expressed in terms of the 
mass processed into stars per comoving volume, and assumed to be
traced primarily by hot young stars.  Knowledge of the SFR is
obtained by analyzing the blue/UV luminosity density using statistical
samples of galaxies.  Blue light is
thought to be a good proxy of star-formation activity, but 
extinction by dust complicates the measurement.

Fig.\ 5 shows various SFRs based on different approaches to the
problem. SFR 4, from Hopkins and Beacom \cite{hb06}, is based on a
compilation of IR, optical, and UV data.  SFR 1 gives the lowest
acceptable rate compatible with the data. SFR 3 gives the rate
assuming large extinction corrections, and was used to fit the UHECR
spectrum in \cite{wda04}. With this rapidly increasing rate density of
GRBs at redshifts $z \approx 1$ -- 3, a large pair production trough
at $\approx 10^{18.4}$ eV is formed \cite{bg88}. A rate density of
GRBs following the extremely active SFR 5, based on IR luminosity
density, is ruled out from calculations of the UHECR spectrum, as can
be seen in Fig.\ 6.

A statistical study \cite{ld07} of the redshift and opening-angle
distribution of GRBs observed before Swift, primarily Beppo-SAX GRBs,
and the $z$-distribution of GRBs observed with Swift, rejects SFR 3
for GRBs based on a standard relativistic jet model.  We concluded
that the GRB rate density must monotonically increase to $z \approx
5$ -- 7 to explain the GRB statistics.  The UHECR spectrum from SFR 6,
seen in Fig.\ 6, gives a reasonable fit to the HiRes data (the SFR 7
spectrum is nearly the same).  Normalization to the Auger data would
change the emissivity normalization down by a factor $\approx 1/3$.

The idealized SFR $\propto (1+z)^4$ sketched in Fig.\ 5 was used by
Berezinsky and his collaborators \cite{bgg06,ber06} to describe the
SFR activity of the UHECR sources, possibly blazars. Such extreme SFR
activity produces a large pair production trough, and they proposed
that the ankle/dip feature in the UHECR spectrum is due to photopair
losses. Reasonable fits to the HiRes data were obtained with injection
indices into intergalactic space $s\approx 2.7$.

This model SFR $\propto (1+z)^4$ can hardly be correct, but models of
blazars are more difficult than of GRBs by requiring both luminosity
and density evolution.  The long-duration GRB engine could very well
be unchanged with epoch, but  both due to black hole growth
and fueling, the UHECR output from blazars must
change with time. This behavior is not
well understood, so a mathematical model might be the best that can be
done pending better studies (cf.\ \cite{der07}).

\begin{figure}
\includegraphics[width=2.8in]{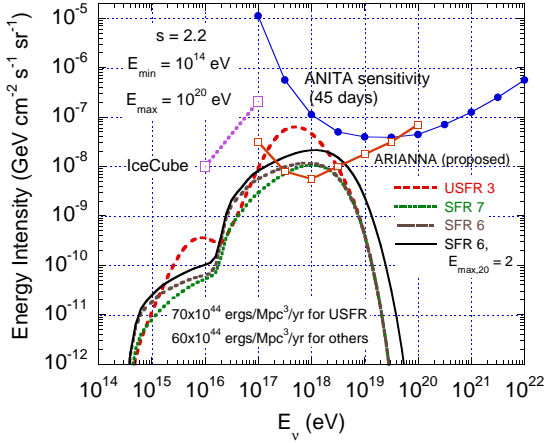} 
\caption{Comparison of model calculations of GZK neutrino intensities
for neutrinos of all flavors with sensitivity curves as labeled.}
\label{f7}
\end{figure}

Figure \ref{f7} shows predictions of the total diffuse neutrino intensity
spectrum for the
various GRB models in Figure 6, compared
with the ANITA sensitivity for a 45 day flight \cite{bar06},
 and an estimated sensitivity for the proposed ARIANNA
project \cite{bar06a}. 
Long-duration balloon-borne high-energy neutino
telescopes can already start to test SFR 3 used in our GRB 
model for UHECRs \cite{wda04}, and 
is close to discovering the guaranteed flux of GZK
neutrinos. The cosmogenic neutrino
flux from GRBs is difficult to detect with IceCube, 
which is more sensitive to cosmic PeV sources 
(see \cite{kar03}
for IceCube sensitivities).

\section{Survival of UHECR Ions in the Blazar Environment}

Spurred by the Auger results on composition \cite{ung07}, we examine
whether UHECRs can leave the blazar environment without being
significantly photo-eroded. This assumes that blazar jets
accelerate UHECRs, which is feasible given that the UHECR 
spectrum can be fit using simple assumptions for the SFR activity and injection spectrum
\cite{bgg06,ber06}. $\gamma$ rays from blazars could be a consequence of 
ultra-relativistic hadrons in blazar jets \cite{mb92,ad03}.  Cosmic rays accelerated
in the inner blazar jet can
power the knots and lobes of FR II radio galaxies through UHE beamed
neutral production and escape due to photopion interactions of UHECR
protons with ambient radiation in the inner jet \cite{ad03}. 
The formation of a neutral beam in FR II radio galaxies 
could explain the Chandra X-ray emission from knots and hot
spots in radio galaxies \cite{hk06} by a second synchrotron component
induced by UHECR activity \cite{ad04}. 

Crucial to efficient neutrino and neutral beam power is that the blazar jet is
found within the BLR where the scattered radiation field is intense.
The question of the location of the $\gamma$-ray production site \cite{jor01,lv03},
which would almost certainly 
be the location of the UHECR accelerator, will be settled by
correlated GLAST/radio observations. 

Ions accelerated in the inner jets of BL Lac objects and FR I radio
galaxies can pass through the broad line region (BLR) without
significant photodisintegration, as we now show.  Underlying this
estimate is the unification scenario for blazars as expressed in
\cite{up95}, in which FR Is are the parent population of BL Lac
objects, and FR IIs are the parent population of FSRQs.

Let the blazar BLR be approximated by a spherically-symmetric
distribution of scattering electrons with density $n_0(x)$ at distance
$x$ from the central source.  The scattered radiation density in
blazars can be estimated by noting that a fraction $\approx
n_0(x)\sigma_{\rm T} (x/2)$ of the ambient photons will be scattered
within a shell of width $\approx x/2$. For an isotropically emitting
central source of radiation with photon production rate $\dot
N_{ph}(\e )$ per unit dimensionless photon energy $\e = h\nu/m_ec^2$,
the ambient photon density from the central source emission is
$n_{ph}(\e;x) = \dot N_{ph}(\e)/4\pi x^2 c$.  For assumed isotropic
Thomson scattering, the spectral density of scattered radiation is
\begin{equation}
n_{sc}(\e;x) \approx {n_0(x)\sigma_{\rm T} \dot N_{ph}(\e )\over 8\pi xc}\;
\label{nsc}
\end{equation}
\cite{dp03,bd95,bl95}.

The central source emission is assumed to be radiated by an accretion
disk around the supermassive black hole. We represent the blue-bump
emission, commonly observed in Seyfert galaxies (it is difficult to
observe in blazars because of the luminous jet radiation), by a
Shakura-Sunyaev disk spectrum of the form
\begin{equation}
\dot N_{ph}(\e ) = L_0 \; 
{\e^{-2/3}\exp(-\e/\e_{max})\over m_ec^2 \e_{max}^{4/3} \Gamma(4/3)}\;,
\label{Nphe}
\end{equation}
normalized to the total Shakura-Sunyaev disk luminosity $L_0 =
10^{46}L_{46}$ ergs s$^{-1}$, where $\Gamma(4/3) \cong 0.893$, and
$\e_{max}$ is the maximum photon energy radiated in the disk spectrum.
For the UV bump observed in supermassive black holes, $\e_{-5} \equiv
\e_{max}/ 10^{-5}\cong 1$.

The blue bump in 3C 273 reaches a $\nu F_\nu$ peak flux of $\approx
3\times 10^{-10}$ ergs cm$^{-2}$ s$^{-1}$ at $\approx 10$ eV,
corresponding to $\nu L_\nu \approx 2\times 10^{46}$ ergs s$^{-1}$ and
$\e_{-5} \cong 2$ \cite{mon97}.  The hard UV emission component
observed from 3C 279 with IUE \cite{pia99} has a peak $\nu L_\nu$
luminosity $\approx 3\times 10^{45}$ ergs s$^{-1}$. Its effective
temperature is $\approx 20,000$ K, corresponding to a mean
dimensionless photon energy $\approx 10^{-5}$.

\begin{figure}[t]
\center
\includegraphics[scale=0.35]{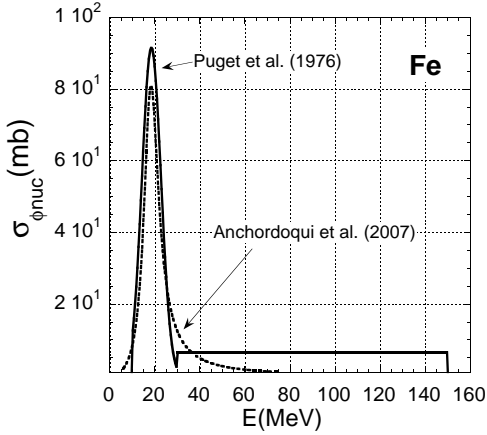}
\caption{Comparison of photo-nuclear destruction cross section 
for Fe
from \cite{psb76} and \cite{anc07}.}
\label{fig08}
\end{figure}

A $\delta$-function approximation for the photodisintegration cross
section of a nucleus with atomic mass $A = 56A_{56}$ is
\begin{equation}
\sigma_A(\e_r ) \cong {\pi\over 2} 
\sigma_{0,A} \Delta \delta(\e_r - \e_{r,0})\;,
\label{sigdelta}
\end{equation}
\cite{anc07},
where $\sigma_{0,A} = 1.45A$ mb, $\Delta \cong 15.6$, $\e_{r,0}
\cong 83.5 A^{0.21}$, and $\e_r$ is the invariant dimensionless photon
energy in the ion's rest frame. A comparison of the photonuclear 
destruction cross sections for Fe is shown in Fig.\ \ref{fig08}.
The $\delta$-function approximation should be fairly good in all 
cases where the target photon spectrum is not too hard.

For an ultra-relativistic ion passing through an isotropic radiation
field, the probability per unit pathlength for the ion to
photodisintegrate by interacting with ambient photons is given by
$${dN_{\phi nuc}\over dx} = (2\gamma^2)^{-1}\times$$
\begin{equation}
 \int_0^\infty d\e\; \e^{-2} n_{ph}(\e )
\int_0^{2\gamma\e } d\e_r \;\e_r\;\sigma_A(\e_r ) \;,
\label{dNscdx}
\end{equation}
where the particle Lorentz factor $\gamma = E/Am_pc^2$.

The probability of an UHECR ion photo-disintegrating as it travels
through the AGN BLR is, using eqs.\ (\ref{nsc}) --
(\ref{sigdelta}) in eq.\ (\ref{dNscdx}), simply $$P_{\phi nuc} \cong
x\;{dN_{\phi nuc}\over dx}\approx $$
\begin{equation}
 {\sigma_{0,A}\Delta \e_{r,0} n_0(x) \sT L_0\over
32 \Gamma(4/3) \g^2 m_e^3 \e_{max}^{3}} \; 
\int_{u_0}^\infty du\; u^{-8/3}\exp(-u) \;,
\label{Phinuc}
\end{equation}
where $u_0 \equiv \e_{r,0}/2\gamma\e_{max}$. 

Here we take the typical extent of the BLR as $\approx 0.1$ -- 1 pc,
and mean optical depth $\tau_{\rm T} \lesssim 0.1$. The BLR medium is
probably clumped in rather dense clouds and has a strong density
gradient from the BLR to the narrow line region \cite{net90,kn99}, but
here we approximate it as being rather uniform within a shell of
radius $10^{18}R_{18}$ cm with Thomson depth $\tau_{\rm T} =
10^{-2}\tau_{-2}$, so that the mean BLR density is $n_0 \cong
1.5\times 10^{4}\tau_{-2}/R_{18}$ cm$^{-3}$.

Eq.\ (\ref{Phinuc}) is easily solved in the limit $u_0 \ll 1$ or $$E
\gg {4\times 10^{15}\over \e_{-5}}\; A^{0.79}\;{\rm eV}\;.$$ The
result is
\begin{equation}
 P_{\phi nuc} \cong  0.12\; {A_{56}^{1.47}
 L_{46}\tau_{-2}\over E_{20}^{1/3} \e_{-5}^{4/3}  R_{18}}\;,
\label{PhinucFe}
\end{equation}
for $E_{20} \gg {10^{-3} A_{56}^{0.79}/ \e_{-5}}$.

This result indicates that for typical parameters that may
characterize the BLR of BL Lac objects, with $\tau_{-2}\lesssim 1$ and
$L_0\approx 10^{44}$ ergs s$^{-1}$, UHECR ions will escape without
undergoing photonuclear breakup. The BLR environment may pose a hazard
to UHECR ion escape in the luminous FR II radio galaxies and FSRQs
with broad optical emission lines. But note that our calculation only
considered a single interaction with the loss of one or a few nucleons
from the UHECR ion. The corresponding probability for complete breakup
will be a factor $\approx A/2$ smaller.

This estimate suggests that UHECR ions can escape from BL Lacs and
also from FSRQs, except in the cases of the most luminous blazars with thick
columns of BLR material.
The scattered radiation field in blazars is a convolution of the
central source luminosity and surrounding gas distribution.  The
$\gamma\gamma$ attenuation process gives a separate probe of this
radiation field \cite{dp03,lb06,rei07}.  By jointly analyzing
photodisintegration and $\gamma\gamma$ processes, GLAST data can be
used to determine if the black hole jet environment limits UHECR escape.

\section{The Extragalactic Background Light}

\begin{figure}[t]
\center
\includegraphics[scale=0.45]{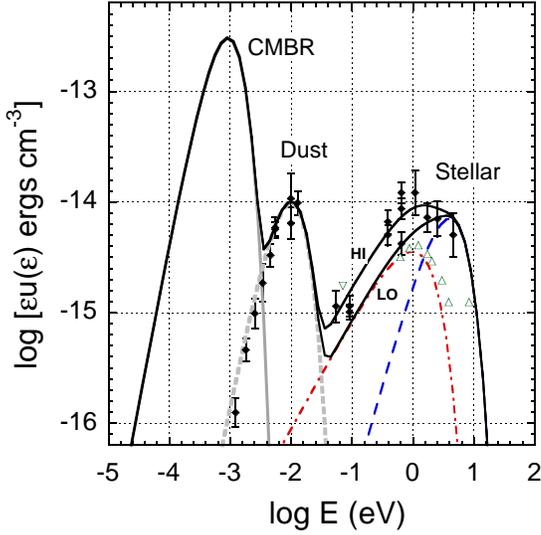}
\caption{Measurements of the EBL at optical and infrared frequencies 
\cite{hd01}, plotted in terms of spectral energy density $\e u(\e )$, 
and fits to the EBL using modified blackbody functions. Data
triangles pointed upwards refer to lower limits, and data triangles
pointed downwards refer to upper limits. The spectral energy density
of the CMBR at $z = 0$ is also shown.}
\label{fig09}
\end{figure}

Fig.\ \ref{fig09} shows measurements of the intensity of the
unresolved IR and optical EBL at the present epoch from the review by
Hauser and Dwek \cite{hd01,sbs07}, including an upper limit at
$\approx 0.1$ eV inferred from $\gamma$-ray observations \cite{sj97}.
Motivated by the appearance of two distinct peaks in the SED of
luminous infrared galaxies \cite{sm96,der97}, by synoptic spectra of
galaxies in the Sloan Digital Sky Survey \cite{obr06}, by SEDs of
nearby normal galaxies, including data from the {\it Spitzer Space
Telescope} \cite{dal07} and normal star-forming galaxies like the
Milky Way \cite{mps06}, I fit these two peaks with modified
blackbody functions.  The lower energy emission
feature peaking near $\approx 0.01$ eV, probably due to radiation
reprocessed by dust, is referred to as the dust component.  The higher
energy emission feature peaking near 2 eV is referred to as the
stellar component. In our calculations, two modified blackbodies make
up the stellar component. More terms can be added as required.

 \begin{table}
\begin{center}
\caption{Properties of the Dust and Two Stellar Components}
\begin{tabular}{lccc}
\hline
Component  & $T$(K) & $u_{0}$  & $k$ \\ 
 &  & ($10^{-14}$ ergs cm$^{-3}$) &  \\ 
\hline
\hline
 Dust & 31 & 0.273 & 3.8  \\
 Star 1 (HI EBL)& 7100 & 1.1& 2.0  \\
 Star 1 (LO EBL)& 7100 & $1.1\div 2$ & 2.0  \\
Star 2 & 16,600 & 0.5 &  3.0 \\
\hline
\end{tabular}
\end{center}
\end{table}

The modified blackbody spectral 
energy density is written in the form
\begin{equation}
\e u(\e ) = u_0\;{w^k\over \exp(w)-1}\;=\; m_ec^2 \e^2 n_{ph}(\e )\;, 
\label{nmodbb}
\end{equation}
where $w \equiv \e/\Theta$. For a blackbody, $k = 4 $ and $u_0 = {8\pi
m_ec^2 \Theta_0^4(1+z)^4/ \lambda_{\rm C}^3} = 6.37\times
10^{-14}(1+z)^4$ ergs cm$^{-3}$.  The fits to the data in Fig.\
\ref{fig09} use the parameters given in Table 2. The high and low
EBLs differ only by a factor $2$ for the intensity of the lower
temperature stellar radiation field.

\begin{figure}[t]
\center
\includegraphics[scale=0.45]{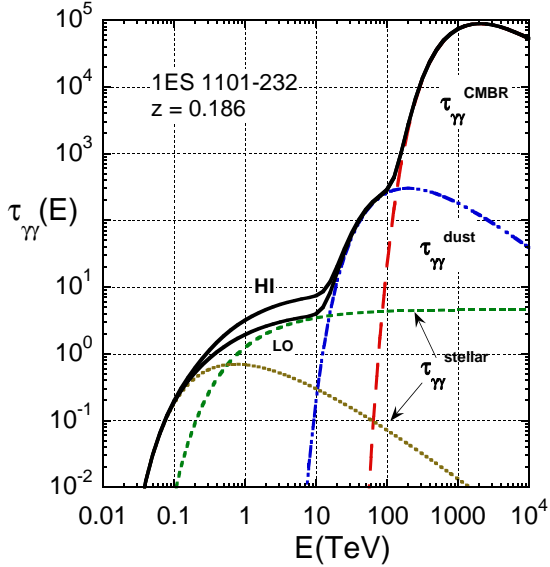}
\caption{Optical depth for a source at $z = 0.186$  to $\g\g$ 
attenuation for the low and high forms for the EBL shown in
Fig.\ \ref{fig09}.}
\label{fig10}
\end{figure}

Fig.\ \ref{fig10} shows the optical depth to $\g\g $ pair production
attenuation for $\gamma$ rays with measured energies $E$ detected 
from the TeV XBL 1ES 1101-232 at $z = 0.186$ \cite{aha07,ben07}.  
Separate components for the CMBR, dust, and
stellar radiation fields are shown for the low EBL in the figure. In
making this calculation, only the CMBR field evolves with redshift,
and the dust and stellar radiation field energy densities remain
roughly constant. This assumption becomes increasingly less reliable
 at higher redshifts.

\begin{figure}[t]
\center
\includegraphics[scale=0.40]{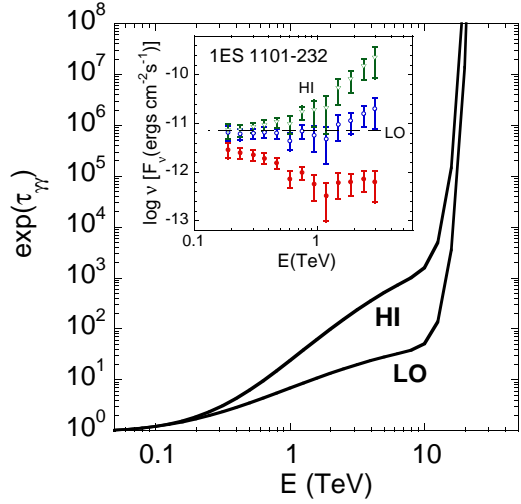}
\caption{Attenuation factor for the low and high forms of the EBL, for 
1ES 1101-232 at $z = 0. 186$. Inset shows the effects of the low and high
EBLs on the intrinsic spectrum of 1ES 1101-232 \cite{aha06,aha07}.}
\label{fig11}
\end{figure}

The attenuation factor, from which the intrinsic spectrum of 1ES
1101-232 is obtained, is plotted in Fig.\ \ref{fig11}.  As can be seen
from the index, the use of the low EBL means that the intrinsic photon
spectral index of 1ES 1101-232 from $\approx 0.2$ -- 3 TeV is $\approx
-2.0$.  If we adopt as a general rule, consistent with our knowledge
of the GeV spectra of FSRQs \cite{muk97} and GeV -- TeV spectra of 
BL Lacs like Mrk
421 and Mrk 501, that the intrinsic spectrum is softer
than $-2$, then the low EBL is favored (cf.\ \cite{mr07,kne04}).  A low EBL between
1 and 10 $\mu$ solves the problem of the unusually hard $\gamma$-ray
spectrum of 1ES 1101-232 \cite{aha06,aha07}, and avoids having to
construct acceleration scenarios not operating in Mrk 421 and Mrk 501, and
to explain lack of evidence of hard synchrotron components associated
with a hard electron component in TeV/XBLs.

The low EBL with a steep 2 -- 10 micron spectrum favored to explain
the TeV blazar data is in general agreement with the Primack model for
galaxy formation \cite{pbs05}, which considers star formation,
supernova feedback and metal production in merging dark matter
halos. The low EBL disagrees with the EBL derived by Stecker and
collaborators \cite{ste07,ms98,ms01}. The problem is that their
empirical data base relies heavily on IRAS data at 12, 25, 60, and 100
$\mu$ and uses a poor representation of the galactic SEDs between 1
and 10 $\mu$. Individual IR and normal galaxies show far more
structure in this region than considered.

\section{The $\gamma$-Ray Horizon}

\begin{figure}[t]
\center
\includegraphics[scale=0.45]{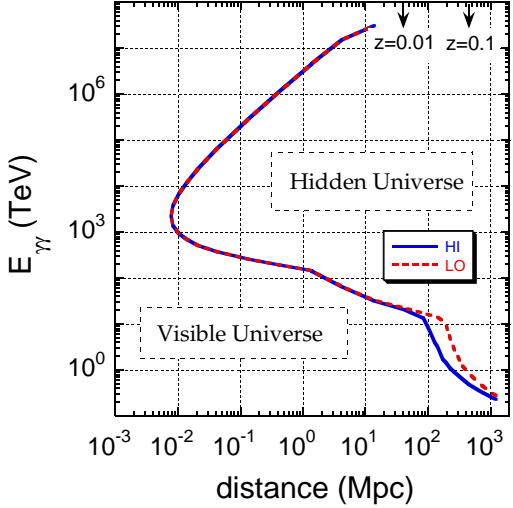}
\caption{Low redshift ($z\ll 1$) $\gamma$-ray horizon giving the 
relationship between photon energy and $z$ or distance where
$\tau_{\g\g} (E,z)= 1$ for the low and high EBLs shown in Fig.\
\ref{fig11}. Calculated up to $z = 0.25$.}
\label{fig12}
\end{figure}

The low and high EBL SEDs represent the likely range of the local
$z\ll 1$ IGM IR and optical radiation fields.  We can use
this field to calculate the photon horizon where $\tau_{\g\g}(E,z) =1$ \cite{gs67,fs70}
in the limit $z \ll 1$, when the IR and optical radiation fields have not
changed appreciably over time. The result is shown in Fig.\
\ref{fig12}. This diagram is primarily illustrative, and in some
respects misleading. At the redshift of 1ES 1101-232, namely $z =
0.186$, this diagram says that the exponential cutoff energy is at
$\approx 300$ -- 400 GeV, and that the low and high EBLs are not
significantly different.  In fact, no exponential cutoff is seen in
the 1ES 1101-232 TeV spectrum (see inset to Figure 11), because the 
actual attenuation is very
sensitive to the full spectrum of the EBL.

This figure does illustrate that PeV $\gamma$-rays can be detected from sources
within our Galaxy, though they might be subject to modest attenuation
from the CMBR. The IR and stellar radiation fields can also contribute 
significant $\g\g$ opacity at $\sim 100$ TeV. Anisotropy effects of the 
radiation fields on opacity have recently been calculated by \cite{mps06,os07}.
Attenuated spectra of TeV -- PeV $\gamma$-ray sources could in principle
give a distance measure of specific sources in the Milky Way and nearby 
galaxies.

\section{The UHECR Ion GZK Radius}

\begin{figure}[t]
\center
\includegraphics[scale=0.36]{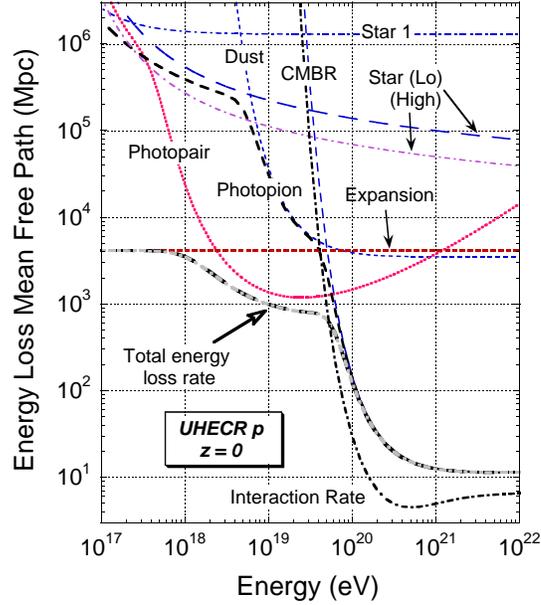}
\caption{Energy loss mean-free paths for protons in the combined
EBL and CMBR radiation fields, including photopion, photopair, and 
expansion losses. The different low and high EBLs make little difference
on the UHECR proton spectrum. }
\label{fig13}
\end{figure}

We now use our low and high estimates of the EBL to calculate the ion
GZK radius.  First we show, in Figure \ref{fig13}, the energy loss and
interaction mean-free paths for UHECR protons interacting with the
combined EBL and CMBR, with components as shown. The dust component
makes a minor, $\simeq 10$\% contribution at $\approx 8\times 10^{19}$
eV, but could be somewhat larger for a modified fit to the FIR
EBL. Other than that, the energy-loss mean free path is
essentially given by the CMBR result (Figure 1).  Note also that the
ratio of the CMBR energy loss and scattering mean-free paths is
$\approx 5$ on the low-energy wing where protons interact with the
exponential Wien portion of the blackbody distribution.  This ratio, 
arising from the 20\% inelasticity for direct and resonance pion production,
decreases at $\gtrsim 4\times 10^{20}$ eV due to the greater fraction
of multipion interactions at higher energies.

Before calculating ion mean-free paths, it is worth mentioning how the
energy-loss formula for photodisintegration is calculated. When a
single nucleon is ejected, then an energy loss $\propto 1/A$ of the
original energy $E$ happens (technically, provided that the nucleon
ejection is isotropic in the nucleon frame), and for the ejection of
two nucleons, an energy loss $= 2E/A$ occurs.  For multi-nucleon
injection, an average factor is used, given by Puget et al.\ (1976)
\cite{psb76}, $= 3.6/A$ for $10 \leq A \leq 22$, and $= 4.349E/A$ for
$23 \leq A \leq 56$.  A low-energy threshold of $10$ MeV is used
\cite{ss99}. Obviously, after a single interaction, the original
nucleonic identification is changed, so that new sets of loss rates
have to be used for the daughter particles. The calculated MFPs have
only a generalized meaning, but provide inputs for accurate Monte
Carlo simulations.

\begin{figure}[t]
\center
\includegraphics[scale=0.36]{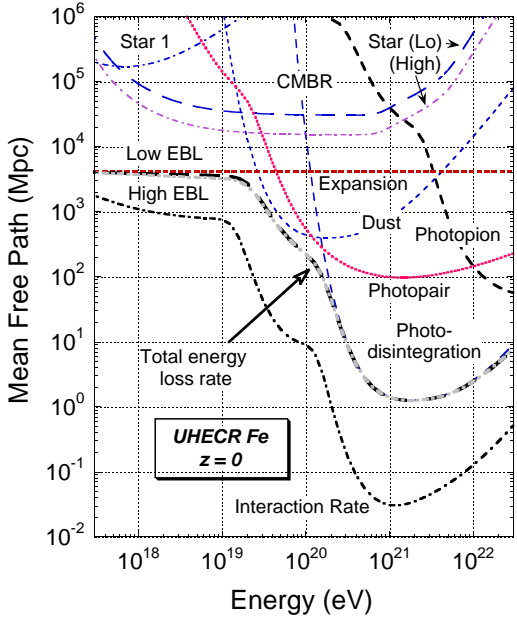}
\caption{Photonuclear MFPs for UHECR Fe, including
separate contributions to photodisintegration, and total energy-loss 
and interaction rates.}
\label{fig14}
\end{figure}

Figure \ref{fig14} gives various contributions of the CMBR and low and
high EBLs to the effective energy-loss rate of UHECR Fe in an IGM
radiation field at the present epoch. Also shown is the
photodisintegration interaction rate for the ejection of at least one
nucleon. In effect, the energy-loss MFP gives the distance an Fe ion
would have to travel to be broken up into mostly protons and
neutrons. For $E \gtrsim 6\times 10^{19}$ eV, Fe only has to travel
${\cal O}($Mpc) before being transformed to lighter elements, and
could hardly be seen in abundance in the UHECRs above this energy unless
UHECR sources reside in the neighborhood of our Galaxy, including M31 
and our satellite galaxies (possible for a GRB origin of the UHECRs).
Secondary nucleons  with $A\sim 56/2$ would 
be more prevalent due to Fe photo-erosion.

\begin{figure}[t]
\center
\includegraphics[scale=0.36]{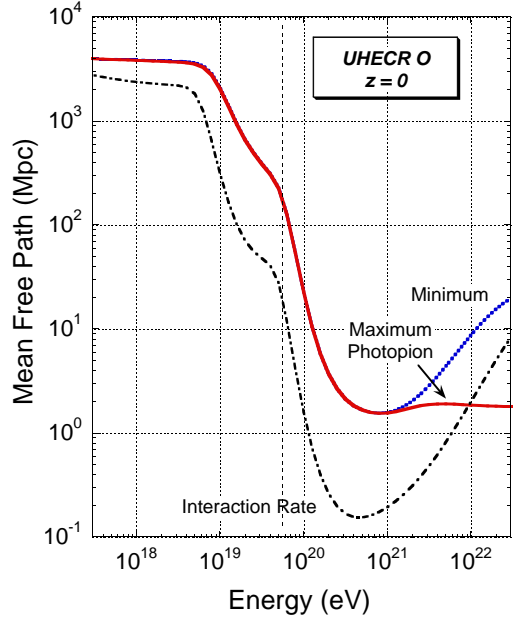}
\caption{Comparison of the effects 
on the total energy-loss MFP of UHECR O in the CMBR and low EBL 
field
for different assumptions of the 
inelasticity in inelastic photopion production.}
\label{fig15}
\end{figure}

Figure 15 compares different assumptions for the photopion energy-loss
rate on the total energy-loss rate of UHECR O in the CMBR and the low
EBL.  Without a detailed physical model, the photopion cross section
should go $\propto A^{2/3}$ for a quasi-spherical nucleon, with a
photopion energy-loss rate $\propto A^{2/3}$, giving the maximum
MFP. More realsitically, only one pion is produced with near threshold
energy in the interaction, so the inelasticity would be $\propto
A^{-1}$, and the photopion energy-loss rate $\propto A^{-1/3}$, giving
the minimum MFP.  In multi-pion production, the inelasticity should be
larger than the minimum, so the ``true" photopion energy loss rate should
reside between these two extremes.  The difference in either case 
is not significant below $10^{21}$ eV, as can be seen.

\begin{figure}[t]
\center
\includegraphics[scale=0.38]{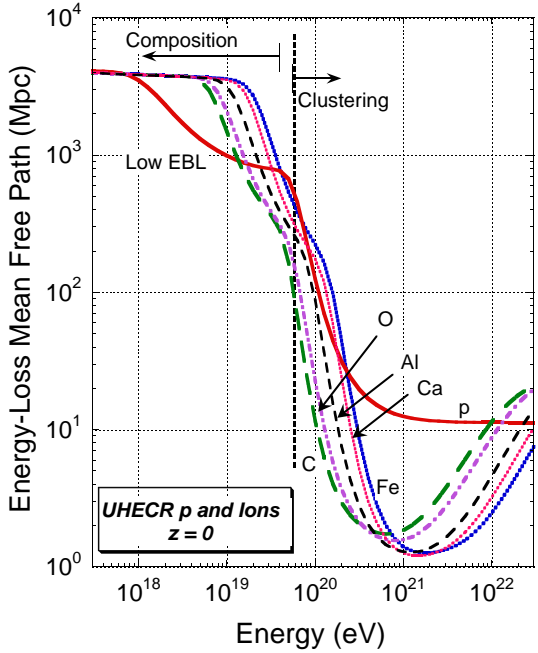}
\caption{Comparison of energy-loss MFPs of UHECR protons and ions
in the combined low EBL and CMBR.}
\label{fig16}
\end{figure}

Proton and ion MFPs for energy loss in the combined CMBR and low
EBL are shown in Figure 16. Most dramatic is the rapid decrease of the
MFP between $10^{19}$ and $10^{20}$ eV, precisely where the HiRes and
Auger Observatories discover the spectral softening in the UHECR
energy spectrum \cite{hires07,yam07}.
Keeping in mind that the essential destruction of a nucleus of a
given type is proportional to the scattering rate, and that the
complete breakup proportional to the energy-loss rate, then this
figure shows that structure within 100s of Mpc become visible above
$\approx 6\times 10^{19}$ eV. At these energies, all ions will have
undergone significant breakup, so that an original enhancement of Fe
would be broken up to smaller $Z$. 
With increased statistics, higher energy events
should be identified with even closer structures.

Figure 16 reveals a number of interesting things. First, 
whereas UHECR protons
have significant photopair losses to $\approx 10^{18}$ eV,
there is
no change in ionic composition below $\approx 5\times 10^{18}$ eV. At these 
energies, all ion losses are due to expansion. Above
$\approx 5\times 10^{18}$ eV, corresponding to the ankle or dip
energy, photopair and photodestruction losses become roughly equally
important.  The dust component of the EBL (whose SED is not that accurately
known) is important for the rapidly increasing energy-loss rate
between $\approx 5\times 10^{18}$ -- $6 \times 10^{19}$ eV.  Above
$\approx 6\times 10^{19}$ eV, photodisintegration by the CMBR starts
to dominate, reducing the interaction length to tens of Mpc or
less. UHECR Fe injected at $\gtrsim 10^{20}$ eV will, in short order,
degrade to lighter nuclei at lower energies, which have larger cross
sections at a given energy to degrade the nuclei.

By the accumulation of lighter $Z$ material below the GZK energy,
 an ankle and cutoff may be formed through the injection
of UHECR ions with large $Z \sim 30$ -- 56. The formation of this dip,
seen in AGASA, HiRes and Auger data, must be explained in any viable
model for UHECRs.  The apparently natural explanation of the ankle as a pair
production trough resulting from UHECR proton injection
 \cite{bg88,wda04,ber06,bgg06}
might also find an explanation in GZK nucleonic physics. Detailed
study requires a Monte Carlo simulation.

\section{IGM Magnetic Field}

We use the Auger results \cite{aug07} 
to set lower limits on the mean IGM magnetic field $B$. 
The equation for the deflection angle from a source at distance
$d$ is \cite{wax95,wc96}
$$\theta_d \simeq {d\over 2r_{\rm L}\sqrt{N_{inv}} } \cong {dZeB\over 2E\sqrt{N_{inv}} }$$
\begin{equation}
\simeq {2.6^\circ }\;\big({Z\over 10}\big)
\;{B_{-11}\;d(100 {\rm ~Mpc})\over E_{20}\sqrt{N_{inv}}}\;.
\label{thetad}
\end{equation}
Here $N_{inv}\cong d/\lambda\gtrsim 1$ is the number of inversions of
the magnetic field, also expressed through the magnetic-field correlation length
$\lambda$.  If the two UHECRs within 3$^\circ$ of Cen A were 
accelerated by the radio jets of Centaurus A, the measured 3$^\circ$ deflection implies that
$$B_{-11} \gtrsim 20\; {(E/6\times 10^{19}{\rm~eV})\over (Z/10)}\;.$$
If the UHECRs originated in fact from the AGNs in the V\'eron-Cetty
and V\'eron catalog \cite{vcv06}, taking $d \cong 75$ Mpc and $\theta_d \cong
3^\circ$ gives $$B_{-11} \gtrsim 0.9\; {(E/6\times
10^{19}{\rm~eV})\over (Z/10)}\;.$$

Both of these values are reasonable, and possibly compatible depending
on gradients in the IGM field. The IGM magnetic field energy density
is a small fraction of the CMBR energy density or the EBL energy
density. This shows the potential power of Auger for measuring IGM
fields \cite{kro01}. If these IGM fields are accurate, $\gamma$-ray
pulse broadening could not be measured \cite{pla95}.

The equation for the time delay due to the deflection of UHECRs from
an impulsive source is
\begin{equation}
\Delta t \cong {d\over 6 c} \theta_d^2 \simeq {350 B_{-11}^2 d^3(10{\rm~Mpc})
\over N_{inv}}\; \big({Z\over 10}\big)^2 
\;{\rm~yr}\;
\label{tauE}
\end{equation}
\cite{wax95,wc96}. Recurrent events over decades or shorter could 
be observed from UHECRs powered by nearby GRBs at the distance of Cen
A or GRB980425/SN1998bw ($d \cong 36$ Mpc) \cite{fol06} in
 favorable circumstances.


\section{Galactic Cosmic Ray Astronomy is Unlikely}

It is great that Auger \cite{aug07} has opened the field of
extragalactic cosmic-ray astronomy, but the Pierre Auger 
Cosmic Ray Observatory 
(and its Northern Hemisphere
counterpart, the Telescope Array, with half of Auger's effective
area) may prove less useful for undertaking cosmic ray astronomy
of galactic sources, simply
because these sources are too weak.

To detect clustering in the arrival directions of signal 
 (the definition of an astronomy), then the source distance 
\begin{equation}
d \ll \sqrt{N_{inv}}r_{\rm L} \cong 10\sqrt{N_{inv}}
\;{E_{20}\over (Z/10)B_{\mu{\rm G}}}\;
{\rm kpc}\;;
\label{dllr}
\end{equation}
otherwise the particle trajectories would be hopelessly scrambled.  To
accelerate particles with energy $E$, equation (\ref{L}), requires
source luminosities $$L \gtrsim {3\times 10^{43}\over (Z/10)^2}\;
E_{20}^2\; {\rm ergs~s}^{-1}\;.$$ Writing this expression as a limit
on particle energy gives, using eq.\ (\ref{dllr}), the maximum source
distance
\begin{equation}
d \;\ll \;20 \;{\sqrt{N_{inv}\;
(L/10^{38}{\rm~ergs~s}^{-1})}\over B_{\mu{\rm G}}}\;{\rm pc}\;
\label{dpc}
\end{equation}
($Z$ drops out of the expression). 
 
In the Milky Way, 
where a large scale ordered field 
is measured  \cite{rma00}, probably not a large number of
inversions could occur over a distance $\approx 100$ pc, 
so I argue that $N_{inv}$ is not large. 
In this case, there are not sufficiently luminous sources of
nonthermal power close to the Solar system that could be observed by  
the cosmic rays accelerated at that source.  
For example, the $\gamma$-ray luminosity of Geminga, $d
\cong 140$ pc, is $\approx 3\times 10^{33}$ ergs s$^{-1}$ \cite{ulm94}.
The shell-type SNR
RX J1713.7-3946, if 1 kpc distant, releases $\approx 10^{34}$ ergs s$^{-1}$
in nonthermal $\gamma$-ray energy and $\lesssim 10\times$ more in nonthermal 
synchrotron radiation \cite{aha06a}. The shell SNR RX J0852.0--4622, with $d\approx 200$ pc, 
releases $\approx 3\times 10^{32}$ ergs s$^{-1}$ \cite{lem07} between 1 and 10 TeV.

This would explain the futile search for Galactic sources of cosmic
rays \cite{abb05,aug07a}, which could only be detected from rare
nearby SNe or hyper-energetic GRBs in our Galaxy \cite{bie04,dh05}.

\section{UHE Source Neutrinos from Blazars and GRBs}

Discussion of neutrino production from 
GRBs will be kept short because
it was recently reviewed
 \cite{da06,mr06}. Our basic
approach is to use measured flare nonthermal 
fluence to get apparent isotropic energy
in the comoving frame within the uncertainty of the Doppler factor.
For GRBs, we use observational constraints, including 
the burst rate and typical energy release, 
to normalize the mean (volume- and time-averaged)
energy emissivity needed to power the UHECRs, from 
which the 
amount of energy that a typical GRB must release in the form of nonthermal
hadrons can be derived. Our results  \cite{wda04} showed that for an origin of UHECRs from 
 long-duration GRBs, with an
upper SFR 3 (Figure 5), long-duration GRB blast waves must be baryon-loaded by a factor
$f_{CR}\gtrsim 60$ compared to the primary electron energy that is
inferred from the $X/\gamma$ GRB flux.

The GRB neutrino fluences are to first order proportional 
to the
electromagnetic $X/\gamma$ radiation fluence from a GRB.  
Neutrino fluences expected in the collapsar GRB scenario from a burst
with photon fluence $\Phi_{rad} = 3\times 10^{-4}\,\rm erg\, cm^{-2}$,
were calculated  \cite{da03}
for 3 values of the Doppler factor $\delta$ from a GRB at redshift $z
= 1$ ($h = 65$). For a GRB collapsar-model
calculation, we assumed that the prompt emission
is contributed by $N_{spk} = 50$ spikes with characteristic timescales
$t_{spk} \simeq 1 \rm \, s$ each, which defines the characteristic
size (in the proper frame) of the emitting region associated with each
individual spike through the relation $R_{spk}^\prime \simeq t_{spk}
\delta /(1+z)$.

The numbers of muon neutrinos for IceCube 
parameters that would be detected from a single collapsar-type GRB
 with a baryon-loading factor
$f_{CR} =20$  for $\delta = 100,\,
200$ and 300 are $N_\nu = 1.32,\, 0.105 $ and 0.016, respectively. For
the large baryon load required for the proposed model of UHECRs, our
calculations showed that 100 TeV -- 100 PeV neutrinos could be
detected several times per year from all GRBs with km-scale
neutrino detectors such as IceCube \cite{da03,wda04}.  Detection of
even 1 or 2 neutrinos from GRBs with IceCube or a northern hemisphere
neutrino detector will provide compelling support for a GRB origin of
UHECRs. See \cite{da06} for more details.

Detailed numerical simulations to calculate neutrino production
in colliding shell scenarios of GRBs, including the diffuse neutrino
intensity, are given in \cite{mn06}. Calculations of neutrino fluxes during
X-ray flares found with Swift was treated in \cite{mn06a}, and
neutrinos and UHECRs from low-luminosity GRBs, in \cite{mur06}.

Following a similar methodology and normalizing to blazar flare
fluences, Armen Atoyan and I \cite{ad01,ad03} calculated 
neutrino fluxes from blazars; see \cite{lev06} for a recent review. We
found that for equal power into ultra-relativistic electrons, which 
makes the 
$X/\gamma$ emission, and protons, IceCube could detect 1 or 2 neutrinos from 
a powerful blazar flare like the 3C 279 flare in 1996
\cite{weh98}, of which GLAST should see one or two per month.

Our principle result was that FSRQs, with strong scattered radiation field,
provide much more effective target photons for photopion production in their
jets
than do BL Lac objects. Thus powerful FSRQs like 3C 279 ($z = 0.54$), 
PKS 0528+134 ($z = 2.06$), CTA 102 ($z = 1.037$),
of 3C 454.3 ($z = 0.86$)
are a better targets
for neutrino telescopes than nearby BL Lac objects. 

It will be important to check these conclusions with new, incoming 
$\gamma$-ray data. The incredibly short flaring timescale 
in the TeV XBL PKS 2155-305
at $z = 0.116$ on time scales of 100s of seconds \cite{ben07},
with  bolometric apparent $\gamma$-ray powers of $\approx 10^{45}$ ergs s$^{-1}$
(depending on EBL assumption), represents a different regime for 
neutrino production in BL Lacs than considered earlier, and a probe
of black holes on small size scales  \cite{bfr07}. 
For blazar modeling of all types, 
GLAST will provide an excellent data base to
determine parameters
used in  models for neutrino production. (see \cite{bec07,ack07}
for blazar neutrino searches.)

\section{Summary}

Auger has already contributed three major discoveries to cosmic-ray 
physics:
\begin{enumerate}
\item the GZK cutoff, also found with HiRes;
\item Mixed ionic composition in the UHECRs up to a $few\times 10^{19}$ eV; and
\item statistical demonstration that the clustering
of 27 UHECRs with  energies $\gtrsim 6\times 10^{19}$ eV  follow the matter
distribution as traced by nearby ($\lesssim 75$ Mpc) AGNs.
\end{enumerate}

These results establish without doubt that UHECRs
originate from astrophysical sources. 
With this information, and based on past theoretical work, 
I argue that GRBs and radio-loud AGNs, which are classified as 
blazars
when viewed on-axis, are the
most probable sources of UHECRs, as can be demonstrated 
by hadronic $\gamma$-ray
signatures. The most convincing evidence would be
direct detection of PeV neutrinos from UHECR sources
with IceCube or KM3NET.
Detection of GZK EeV neutrinos from photopion interactions of UHECRs
as they propagate through EBL with an Askaryan telescope like ANITA will
also importantly test astrophysical models.

In this paper, I used $\gamma$-ray observations to constrain the 
spectrum of the EBL between 1 and 100 $\mu$, where it is poorly 
known. By adopting as a general rule, consistent with observations,
that blazar $\gtrsim 100$ MeV -- TeV $\gamma$-ray spectra are steeper
than $-2$, a low EBL, shown in Figure 9, was favored. With this
EBL, we can deconvolve the intrinsic source spectrum of low-redshift sources, and
calculate the photodisintegration rate of different UHECR ions. 
The GZK curves for different ionic species obtained with the low
EBL is shown in Figure 16. These results are in accord with the location
of the GZK spectral cutoff measured with Auger and HiRes, 
and a mixed ionic composition, because
protons are more difficult to accelerate, and Fe will be broken up. 

GRBs and blazars are both viable sites for UHECR
acceleration, consistent with the Auger results. 
What will be important is to associate UHECR arrival directions
with a subset of AGNs or galaxies. Small metal-poor bluish 
star-forming galaxies are likely hosts of long-duration GRBs \cite{flo03}.
Radio galaxies would host misaligned blazars, and slightly misaligned
blazars could be $\gamma$-ray dim and cosmic-ray bright. GLAST will
be able to see $\gamma$-ray dim AGNs (the radio galaxies M87 \cite{bei07}, Cen A
\cite{har99}, and 
NGC 6251 \cite{muk02}
are definite, likely, and probable $\gamma$-ray sources, respectively).
Associating radio galaxies 
or GLAST $\gamma$-ray galaxies in the nearby universe with arrival directions
of UHECRs could decide the question of blazar origin of the UHECRs. 
With the $\gamma$-ray, cosmic ray, and neutrino observations,
it is likely that the problem of  UHECR origin will
soon be solved\footnote{The theoretical basis 
for the results presented here are given in my book
with Govind Menon entitled ``High-Energy Radiation from Black Holes:
$\gamma$-rays, cosmic rays, and neutrinos," to be published 
by Princeton University Press in 2008.}.


\vskip0.2in

\section{Acknowledgements}

I would like to thank Justin Finke, Soebur Razzaque, David Seckel,
and Stefan Wagner for discussions, Guido Barbiellini
for a very enjoyable talk over coffee, 
Alan Watson for correspondence, Armen Atoyan for
collaboration, and Jeremy Holmes for his skillful coding. I would also
like to thank Xiang-Yu Wang, Soebur Razzaque, and Peter M\'esz\'aros
for informing me of their work prior to submission, and for a
useful discussion on these issues.  I am grateful to Magda Gonz\'alez
for calculating the BATSE yearly-average fluence from GRBs.  I would
finally like to thank the organizers, especially Alberto Carrami\~nana
and Gustavo Medina-Tanco, for their kind invitation to the M\'erida
Yucatan ICRC.  This work is supported by the Office of Naval Research,  NASA
{\it GLAST} Science Investigation DPR-S-1563-Y, and  NASA {\it
Swift} Guest Investigator Grant DPR-NNG05ED411.

\bibliography{icrc1160_final}

\begin{thebibliography}{100}

\bibitem{hires07}
{HiRes Collaboration}, ArXiv Astrophysics e-prints {\bf 0703099},    (2007).

\bibitem{yam07}
T. {Yamamoto} and {for the Pierre Auger Collaboration}, ArXiv e-prints {\bf
  707},    (2007).

\bibitem{gre66}
K. {Greisen}, Physical Review Letters {\bf 16},  748  (1966).

\bibitem{zk66}
G.~T. {Zatsepin} and V.~A. {Kuz'min}, Soviet Journal of Experimental and
  Theoretical Physics Letters {\bf 4},  78  (1966).

\bibitem{ung07}
M. {Unger} and {for the Pierre Auger Collaboration}, ArXiv e-prints {\bf 706},
    (2007).

\bibitem{nw00}
M. {Nagano} and A.~A. {Watson}, Reviews of Modern Physics {\bf 72},  689
  (2000).

\bibitem{ta04}
D.~F. {Torres} and L.~A. {Anchordoqui}, Reports of Progress in Physics {\bf
  67},  1663  (2004).

\bibitem{alo07}
R. {Aloisio} {\it et~al.}, Astroparticle Physics {\bf 27},  76  (2007).

\bibitem{ad01}
A. {Atoyan} and C.~D. {Dermer}, Physical Review Letters {\bf 87},  221102
  (2001).

\bibitem{sta00}
T. {Stanev} {\it et~al.}, \prd {\bf 62},  093005  (2000).

\bibitem{czs92}
M.~J. {Chodorowski}, A.~A. {Zdziarski}, and M. {Sikora}, \apj {\bf 400},  181
  (1992).

\bibitem{psb76}
J.~L. {Puget}, F.~W. {Stecker}, and J.~H. {Bredekamp}, \apj {\bf 205},  638
  (1976).

\bibitem{ss99}
F.~W. {Stecker} and M.~H. {Salamon}, \apj {\bf 512},  521  (1999).

\bibitem{bir95}
D.~J. {Bird} {\it et~al.}, \apj {\bf 441},  144  (1995).

\bibitem{aug07}
{The Pierre Auger Collaboration}, Science {\bf 319},  939  (2007).

\bibitem{sta95}
T. {Stanev} {\it et~al.}, Physical Review Letters {\bf 75},  3056  (1995).

\bibitem{tor03}
D.~F. {Torres} {\it et~al.}, \physrep {\bf 382},  303  (2003).

\bibitem{bg99}
E. {Boldt} and P. {Ghosh}, \mnras {\bf 307},  491  (1999).

\bibitem{hal07}
F. {Halzen}, ArXiv e-prints {\bf 710},    (2007).

\bibitem{ave04}
M. {Ave} {\it et~al.}, Nuclear Physics B Proceedings Supplements {\bf 136},
  159  (2004).

\bibitem{anc07a}
L.~A. {Anchordoqui}, D. {Hooper}, S. {Sarkar}, and A.~M. {Taylor}, Astropart.\
  Phys. {\bf 23},  11  (2007).

\bibitem{hil84}
A.~M. {Hillas}, \araa {\bf 22},  425  (1984).

\bibitem{abb05}
R.~U. {Abbasi} {\it et~al.}, \apj {\bf 623},  164  (2005).

\bibitem{vie95}
M. {Vietri}, \apj {\bf 453},  883  (1995).

\bibitem{wax95}
E. {Waxman}, Physical Review Letters {\bf 75},  386  (1995).

\bibitem{pre00}
R.~D. {Preece} {\it et~al.}, \apjs {\bf 126},  19  (2000).

\bibitem{ban02}
D.~L. {Band}, \apj {\bf 578},  806  (2002).

\bibitem{ste00}
F.~W. {Stecker}, Astroparticle Physics {\bf 14},  207  (2000).

\bibitem{lia07}
E. {Liang}, B. {Zhang}, F. {Virgili}, and Z.~G. {Dai}, \apj {\bf 662},  1111
  (2007).

\bibitem{wan07}
X.-Y. {Wang}, S. {Razzaque}, P. {M{\'e}sz{\'a}ros}, and Z.-G. {Dai}, \prd {\bf
  76},  083009  (2007).

\bibitem{wda04}
S.~D. {Wick}, C.~D. {Dermer}, and A. {Atoyan}, Astroparticle Physics {\bf 21},
  125  (2004).

\bibitem{sre98}
P. {Sreekumar} {\it et~al.}, \apj {\bf 494},  523  (1998).

\bibitem{smr04}
A.~W. {Strong}, I.~V. {Moskalenko}, and O. {Reimer}, \apj {\bf 613},  956
  (2004).

\bibitem{der07}
C.~D. {Dermer}, \apj {\bf 659},  958  (2007).

\bibitem{kd99}
J.~G. {Kirk} and P. {Duffy}, Journal of Physics G Nuclear Physics {\bf 25},
  163  (1999).

\bibitem{ga99}
Y.~A. {Gallant} and A. {Achterberg}, \mnras {\bf 305},  L6  (1999).

\bibitem{gak99}
Y.~A. {Gallant}, A. {Achterberg}, and J.~G. {Kirk}, \aaps {\bf 138},  549
  (1999).

\bibitem{wax04}
E. {Waxman}, New Journal of Physics {\bf 6},  140  (2004).

\bibitem{dh01}
C.~D. {Dermer} and M. {Humi}, \apj {\bf 556},  479  (2001).

\bibitem{mr92}
M.~J. {Rees} and P. {M\'esz\'aros}, \mnras {\bf 258},  41P  (1992).

\bibitem{vie98}
M. {Vietri}, \apj {\bf 507},  40  (1998).

\bibitem{vc94}
R.~C. {Vermeulen} and M.~H. {Cohen}, \apj {\bf 430},  467  (1994).

\bibitem{mgc92}
L. {Maraschi}, G. {Ghisellini}, and A. {Celotti}, \apjl {\bf 397},  L5  (1992).

\bibitem{muk97}
R. {Mukherjee} {\it et~al.}, \apj {\bf 490},  116  (1997).

\bibitem{kca02}
H. {Krawczynski}, P.~S. {Coppi}, and F. {Aharonian}, \mnras {\bf 336},  721
  (2002).

\bibitem{jor05}
S.~G. {Jorstad} {\it et~al.}, \aj {\bf 130},  1418  (2005).

\bibitem{aha07a}
F. {Aharonian} {\it et~al.}, \apjl {\bf 664},  L71  (2007).

\bibitem{wrm07}
X.-Y. {Wang}, S. {Razzaque}, and P. {Meszaros}, ArXiv e-prints {\bf 0711},
  (2007).

\bibitem{bel03}
A.~M. {Beloborodov}, \apj {\bf 588},  931  (2003).

\bibitem{dcb99}
C.~D. {Dermer}, J. {Chiang}, and M. {B{\"o}ttcher}, \apj {\bf 513},  656
  (1999).

\bibitem{mw01}
P. {M{\'e}sz{\'a}ros} and E. {Waxman}, Physical Review Letters {\bf 87},
  171102  (2001).

\bibitem{fb05}
A.~S. {Friedman} and J.~S. {Bloom}, \apj {\bf 627},  1  (2005).

\bibitem{sod04}
A.~M. {Soderberg} {\it et~al.}, \nat {\bf 430},  648  (2004).

\bibitem{har99}
R.~C. {Hartman} {\it et~al.}, \apjs {\bf 123},  79  (1999).

\bibitem{bbp97}
V.~S. {Berezinsky}, P. {Blasi}, and V.~S. {Ptuskin}, \apj {\bf 487},  529
  (1997).

\bibitem{ino07}
S. {Inoue}, G. {Sigl}, F. {Miniati}, and E. {Armengaud}, ArXiv Astrophysics
  e-prints  (2007).

\bibitem{bd05}
R.~C. {Berrington} and C.~D. {Dermer}, ArXiv Astrophysics e-prints  (2004).

\bibitem{ino07a}
S. {Inoue}, G. {Sigl}, F. {Miniati}, and E. {Armengaud}, ArXiv e-prints {\bf
  711},    (2007).

\bibitem{bd03}
R.~C. {Berrington} and C.~D. {Dermer}, \apj {\bf 594},  709  (2003).

\bibitem{gb03}
S. {Gabici} and P. {Blasi}, \apj {\bf 583},  695  (2003).

\bibitem{pee07}
A. {Pe'er} {\it et~al.}, \apjl {\bf 664},  L1  (2007).

\bibitem{hur94}
K. {Hurley} {\it et~al.}, \nat {\bf 372},  652  (1994).

\bibitem{bd98}
M. {Bottcher} and C.~D. {Dermer}, \apjl {\bf 499},  L131  (1998).

\bibitem{som94}
M. {Sommer} {\it et~al.}, \apjl {\bf 422},  L63  (1994).

\bibitem{gon03}
M.~M. {Gonz{\'a}lez} {\it et~al.}, \nat {\bf 424},  749  (2003).

\bibitem{pw04}
A. {Pe'er} and E. {Waxman}, \apjl {\bf 603},  L1  (2004).

\bibitem{da04}
C.~D. {Dermer} and A. {Atoyan}, \aap {\bf 418},  L5  (2004).

\bibitem{ad03}
A.~M. {Atoyan} and C.~D. {Dermer}, \apj {\bf 586},  79  (2003).

\bibitem{der07a}
C.~D. {Dermer}, \apj {\bf 664},  384  (2007).

\bibitem{tag05}
G. {Tagliaferri} {\it et~al.}, \nat {\bf 436},  985  (2005).

\bibitem{rm98}
J.~P. {Rachen} and P. {M{\'e}sz{\'a}ros}, \prd {\bf 58},  123005  (1998).

\bibitem{zha06}
B. {Zhang} {\it et~al.}, \apj {\bf 642},  354  (2006).

\bibitem{kra04}
H. {Krawczynski} {\it et~al.}, \apj {\bf 601},  151  (2004).

\bibitem{bot05}
M. {B{\"o}ttcher}, \apj {\bf 621},  176  (2005).

\bibitem{brp05}
A. {Reimer}, M. {B{\"o}ttcher}, and S. {Postnikov}, \apj {\bf 630},  186
  (2005).

\bibitem{rmz04}
S. {Razzaque}, P. {M{\'e}sz{\'a}ros}, and B. {Zhang}, \apj {\bf 613},  1072
  (2004).

\bibitem{drl07}
C.~D. {Dermer}, E. {Ramirez-Ruiz}, and T. {Le}, \apjl {\bf 664},  L67  (2007).

\bibitem{weh98}
A.~E. {Wehrle} {\it et~al.}, \apj {\bf 497},  178  (1998).

\bibitem{har01}
R.~C. {Hartman} {\it et~al.}, \apj {\bf 553},  683  (2001).

\bibitem{tes07}
M. {Teshima} {\it et~al.}, ArXiv e-prints {\bf 709},    (2007).

\bibitem{der02}
C.~D. {Dermer}, \apj {\bf 574},  65  (2002).

\bibitem{hb06}
A.~M. {Hopkins} and J.~F. {Beacom}, \apj {\bf 651},  142  (2006).

\bibitem{bg88}
V.~S. {Berezinskii} and S.~I. {Grigor'eva}, \aap {\bf 199},  1  (1988).

\bibitem{ld07}
T. {Le} and C.~D. {Dermer}, \apj {\bf 661},  394  (2007).

\bibitem{bgg06}
V. {Berezinsky}, A. {Gazizov}, and S. {Grigorieva}, \prd {\bf 74},  043005
  (2006).

\bibitem{ber06}
V. {Berezinsky}, Journal of Physics Conference Series {\bf 47},  142  (2006).

\bibitem{bar06}
S.~W. {Barwick} {\it et~al.}, Physical Review Letters {\bf 96},  171101
  (2006).

\bibitem{bar06a}
S.~W. {Barwick}, ArXiv Astrophysics e-prints  (2006).

\bibitem{kar03}
A. {Karle} {\it et~al.}, Nuclear Physics B Proceedings Supplements {\bf 118},
  388  (2003).

\bibitem{mb92}
K. {Mannheim} and P.~L. {Biermann}, \aap {\bf 253},  L21  (1992).

\bibitem{hk06}
D.~E. {Harris} and H. {Krawczynski}, \araa {\bf 44},  463  (2006).

\bibitem{ad04}
A. {Atoyan} and C.~D. {Dermer}, \apj {\bf 613},  151  (2004).

\bibitem{jor01}
S.~G. {Jorstad} {\it et~al.}, \apj {\bf 556},  738  (2001).

\bibitem{lv03}
A. {L{\"a}hteenm{\"a}ki} and E. {Valtaoja}, \apj {\bf 590},  95  (2003).

\bibitem{up95}
C.~M. {Urry} and P. {Padovani}, \pasp {\bf 107},  803  (1995).

\bibitem{dp03}
A.-C. {Donea} and R.~J. {Protheroe}, Astroparticle Physics {\bf 18},  377
  (2003).

\bibitem{bd95}
M. {Boettcher} and C.~D. {Dermer}, \aap {\bf 302},  37  (1995).

\bibitem{bl95}
R.~D. {Blandford} and A. {Levinson}, \apj {\bf 441},  79  (1995).

\bibitem{mon97}
C. {von Montigny} {\it et~al.}, \apj {\bf 483},  161  (1997).

\bibitem{pia99}
E. {Pian} {\it et~al.}, \apj {\bf 521},  112  (1999).

\bibitem{anc07}
L.~A. {Anchordoqui} {\it et~al.}, \prd {\bf 75},  063001  (2007).

\bibitem{net90}
H. {Netzer} {\it et~al.}, {\em {Active Galactic Nuclei}} (Saas-Fee Advanced
  Course 20.~Lecture Notes 1990.~Swiss Society for Astrophysics and Astronomy,
  XII, 280 pp.~97 figs..~ Springer-Verlag Berlin Heidelberg New York, Saas-Fee,
  1990).

\bibitem{kn99}
S. {Kaspi} and H. {Netzer}, \apj {\bf 524},  71  (1999).

\bibitem{lb06}
H.~T. {Liu} and J.~M. {Bai}, \apj {\bf 653},  1089  (2006).

\bibitem{rei07}
A. {Reimer}, \apj {\bf 665},  1023  (2007).

\bibitem{hd01}
M.~G. {Hauser} and E. {Dwek}, \araa {\bf 39},  249  (2001).

\bibitem{sbs07}
F.~W. {Stecker}, M.~G. {Baring}, and E.~J. {Summerlin}, \apjl {\bf 667},  L29
  (2007).

\bibitem{sj97}
F.~W. {Stecker} and O.~C. {de Jager}, \apj {\bf 476},  712  (1997).

\bibitem{sm96}
D.~B. {Sanders} and I.~F. {Mirabel}, \araa {\bf 34},  749  (1996).

\bibitem{der97}
C.~D. {Dermer}, J. {Bland-Hawthorn}, J. {Chiang}, and K. {McNaron-Brown}, \apjl
  {\bf 484},  L121  (1997).

\bibitem{obr06}
M. {Obri{\'c}} {\it et~al.}, \mnras {\bf 370},  1677  (2006).

\bibitem{dal07}
D.~A. {Dale} {\it et~al.}, \apj {\bf 655},  863  (2007).

\bibitem{mps06}
I.~V. {Moskalenko}, T.~A. {Porter}, and A.~W. {Strong}, \apjl {\bf 640},  L155
  (2006).

\bibitem{aha07}
F. {Aharonian} {\it et~al.}, \aap {\bf 470},  475  (2007).

\bibitem{ben07}
W. {Benbow} {\it et~al.}, ArXiv e-prints {\bf 709},    (2007).

\bibitem{mr07}
D. {Mazin} and M. {Raue}, \aap {\bf 471},  439  (2007).

\bibitem{kne04}
T.~M. {Kneiske}, T. {Bretz}, K. {Mannheim}, and D.~H. {Hartmann}, \aap {\bf
  413},  807  (2004).

\bibitem{aha06}
F. {Aharonian} {\it et~al.}, \nat {\bf 440},  1018  (2006).

\bibitem{pbs05}
J.~R. {Primack}, J.~S. {Bullock}, and R.~S. {Somerville},  in {\em High Energy
  Gamma-Ray Astronomy}, Vol.~745 of {\em American Institute of Physics
  Conference Series}, edited by F.~A. {Aharonian}, H.~J. {V{\"o}lk}, and D.
  {Horns} (AIP, New York, 2005), pp.\ 23--33.

\bibitem{ste07}
F.~W. {Stecker}, ArXiv e-prints {\bf 709},    (2007).

\bibitem{ms98}
M.~A. {Malkan} and F.~W. {Stecker}, \apj {\bf 496},  13  (1998).

\bibitem{ms01}
M.~A. {Malkan} and F.~W. {Stecker}, \apj {\bf 555},  641  (2001).

\bibitem{gs67}
R.~J. {Gould} and G.~P. {Schr{\'e}der}, Physical Review {\bf 155},  1408
  (1967).

\bibitem{fs70}
G.~G. {Fazio} and F.~W. {Stecker}, \nat {\bf 226},  135  (1970).

\bibitem{os07}
E. {Orlando} and A.~W. {Strong}, \apss {\bf 309},  359  (2007).

\bibitem{wc96}
E. {Waxman} and P. {Coppi}, \apjl {\bf 464},  L75  (1996).

\bibitem{vcv06}
M.-P. {V{\'e}ron-Cetty} and P. {V{\'e}ron}, \aap {\bf 455},  773  (2006).

\bibitem{kro01}
P.~P. {Kronberg},  in {\em American Institute of Physics Conference Series},
  Vol.~558 of {\em American Institute of Physics Conference Series}, edited by
  F.~A. {Aharonian} and H.~J. {V{\"o}lk} (AIP, New York, 2001), pp.\ 451--462.

\bibitem{pla95}
R. {Plaga}, \nat {\bf 374},  430  (1995).

\bibitem{fol06}
S. {Foley} {\it et~al.}, \aap {\bf 447},  891  (2006).

\bibitem{rma00}
M.~J. {Reid}, K.~M. {Menten}, and A.~L. {Argon}, The Origins of Galactic
  Magnetic Fields, 24th meeting of the IAU, Joint Discussion 14, August 2000,
  Manchester, England, meeting abstract. {\bf 14},    (2000).

\bibitem{ulm94}
M.~P. {Ulmer}, \apjs {\bf 90},  789  (1994).

\bibitem{aha06a}
F. {Aharonian} {\it et~al.}, \aap {\bf 449},  223  (2006).

\bibitem{lem07}
{H.~E.~S.~S.~Collaboration: M.~Lemoine-Goumard} {\it et~al.}, ArXiv e-prints
  {\bf 709},    (2007).

\bibitem{aug07a}
{The Pierre AUGER Collaboration}, Astroparticle Physics {\bf 27},  244  (2007).

\bibitem{bie04}
P.~L. {Biermann}, G. {Medina Tanco}, R. {Engel}, and G. {Pugliese}, \apjl {\bf
  604},  L29  (2004).

\bibitem{dh05}
C.~D. {Dermer} and J.~M. {Holmes}, \apjl {\bf 628},  L21  (2005).

\bibitem{da06}
C.~D. {Dermer} and A. {Atoyan}, New Journal of Physics {\bf 8},  122  (2006).

\bibitem{mr06}
P. {M\'esz\'aros} and S. {Razzaque}, ArXiv Astrophysics e-prints  (2006).

\bibitem{da03}
C.~D. {Dermer} and A. {Atoyan}, Physical Review Letters {\bf 91},  071102
  (2003).

\bibitem{mn06}
K. {Murase} and S. {Nagataki}, \prd {\bf 73},  063002  (2006).

\bibitem{mn06a}
K. {Murase} and S. {Nagataki}, Physical Review Letters {\bf 97},  051101
  (2006).

\bibitem{mur06}
K. {Murase}, K. {Ioka}, S. {Nagataki}, and T. {Nakamura}, \apjl {\bf 651},  L5
  (2006).

\bibitem{lev06}
A. {Levinson}, International Journal of Modern Physics A {\bf 21},  6015
  (2006).

\bibitem{bfr07}
M.~C. {Begelman}, A.~C. {Fabian}, and M.~J. {Rees}, ArXiv e-prints {\bf 709},
   (2007).

\bibitem{bec07}
J.~K. {Becker} {\it et~al.}, Astroparticle Physics {\bf 28},  98  (2007).

\bibitem{ack07}
M. {Ackermann}, \apss {\bf 309},  421  (2007).

\bibitem{flo03}
E. {Le Floc'h} {\it et~al.}, \aap {\bf 400},  499  (2003).

\bibitem{bei07}
M. {Beilicke} {\it et~al.},  in {\em American Institute of Physics Conference
  Series}, Vol.~921 of {\em American Institute of Physics Conference Series},
  edited by S. {Ritz}, P. {Michelson}, and C.~A. {Meegan} (AIP, New York,
  2007), pp.\ 147--149.

\bibitem{muk02}
R. {Mukherjee}, J. {Halpern}, N. {Mirabal}, and E.~V. {Gotthelf}, \apj {\bf
  574},  693  (2002).

\end{thebibliography}
\bibliographystyle{prsty}
\end{document}